\documentclass[twocolumn]{aastex61}
\received{March 8, 2017}
\revised{Feb 9, 2018}
\accepted{Feb 15, 2018}
\submitjournal{ApJ}
\shorttitle{A WISE Survey of New Star Clusters}
\shortauthors{Ryu \& Lee}

\begin{document}

\title{A WISE Survey of New Star Clusters in the Central Plane Region of the Milky Way}

\correspondingauthor{Myung Gyoon Lee}
\email{mglee@astro.snu.ac.kr}

\author{Jinhyuk Ryu}
\affil{Astronomy Program, Department of Physics and Astronomy, Seoul National University, Korea}
\email{ryujh@astro.snu.ac.kr}

\author{Myung Gyoon Lee}
\affiliation{Astronomy Program, Department of Physics and Astronomy, Seoul National University, Korea}

\begin{abstract}
We present the discovery of new star clusters in the central plane region ($|l|<30\arcdeg$ and $|b|<6\arcdeg$) of the Milky Way. In order to overcome the extinction problem and the spatial limit of previous surveys, we use the Wide-field Infrared Survey Explorer (WISE) data to find clusters. We also use other infrared survey data in the archive for additional analysis. We find 923 new clusters, of which 202 clusters are embedded clusters. These clusters are concentrated toward the Galactic plane and show a symmetric distribution with respect to the Galactic latitude. The embedded clusters show a stronger concentration to the Galactic plane than the non-embedded clusters. The new clusters are found more in the first Galactic quadrant, while previously known clusters are found more in the fourth Galactic quadrant. The spatial distribution of the combined sample of known clusters and new clusters is approximately symmetric with respect to the Galactic longitude. We estimate reddenings, distances, and relative ages of the 15 class A clusters using theoretical isochrones. Ten of them are relatively old (age $>800$ Myr) and five are young (age $\approx4$ Myr).
\end{abstract}


\section{Introduction}
Star clusters with known distances are considered to be a good tracer of galactic structure.
We can map the structure of the Galactic disk and the locations of the spiral arms using the distances to young star clusters. However, we have relied on indirect distance estimating methods such as counting the numbers of stars \citep[e.g.][]{str03} or measuring velocities of molecular clouds \citep[e.g.][]{dam01} to describe the structure of distant sites of the Galactic disk. These are because most open clusters with known distances ($\sim2000$, $94\%$ of the open clusters in the catalogs; \citealt{dia02}) are located in the solar neighborhood; $78\%$ of the known-distance open clusters are located closer than 3 kpc from the Sun. This means that it is still difficult to find clusters beyond the solar neighborhood and estimate their distances.


It is expected that a large number of young star clusters are located in the direction of the central Galactic plane region. There are several major disk structures in this direction: the Sagittarius arm, the Scutum-Centaurus arm, the Norma arm, and the two endpoints of the Galactic bar. All of these structures are closely related to the formation of stars and clusters. However, only a tip of the entire cluster population can be seen in optical surveys, because they are often associated with interstellar dust and gas. Thus, a large population of the star clusters in the direction of the central Galactic plane region remains to be discovered.

In order to overcome the extinction problem in optical data, Near-Infrared (NIR) data are often used because they suffer much less from the extinction problem than optical data. Among wide-field NIR data, the Two Micron All Sky Survey (2MASS; \citealt{skr06}) is the most widely used for NIR surveys of star clusters. The great advantage of the 2MASS is that it covers the entire sky. A number of previous studies using the 2MASS data found more than one thousand clusters and cluster candidates around the entire Galactic plane region \citep{iva02, bic03b, dut03, kro06, fro07}. However, these clusters are limited only to the relatively bright sample due to the shallow magnitude limit of the 2MASS.

There are two ways to find the faint clusters that could not be found using the 2MASS data. The first is to use NIR survey data deeper than the 2MASS, such as the UKIRT Infrared Deep Sky Survey (UKIDSS; \citealt{law07}) and the VISTA Variables in the Via Lactea Survey (VVV; \citealt{min10, sai12}). Several studies using these deep NIR data have found several hundred clusters in total \citep{bor11, sol12, sol14, bor14, bar15}. The second is to use data with longer wavelengths than the 2MASS $K_s$ band. This wavelength regime is less affected by interstellar extinction than the 2MASS bands. For example, \citet{mer05} and \citet{mor13} used the \textit{Spitzer Space Telescope} Galactic Legacy Infrared Mid-Plane Survey Extraordinaire (GLIMPSE; \citealt{ben03, chu09}) data, which were obtained from images of $3.6\mu\mathrm{m}$, $4.5\mu\mathrm{m}$, $5.8\mu\mathrm{m}$, and $8.0\mu\mathrm{m}$ filters. They found over one hundred clusters in the Milky Way. However, all of those surveys covered only a part of the Galactic plane. 

In this study, we try to find new clusters in the wide area of the central Galactic plane region following the second way. For this purpose, we use the Wide-field Infrared Survey Explorer (WISE) data \citep{wri10}. The wavelength and all-sky coverage of the WISE data allow us to find clusters hidden deep inside the Galactic plane. During this survey, several studies \citep{maj13, cam15a, cam15b, cam15c, cam16} have reported the discovery of new star clusters and candidates in the Milky Way using the WISE data. However, their findings are mostly located in the second and third Galactic quadrants. Our survey focuses on the Galactic plane region in the first and fourth Galactic quadrants, which have higher extinctions and higher stellar number densities than the second and third Galactic quadrants. We also utilize other infrared data in the archive: the 2MASS, the UKIDSS, the VVV, and the GLIMPSE data.

This paper is organized as follows. In Section 2, we describe the data we used. The search method and the classification of our selected targets are explained in Section 3. We present a list of new clusters found in this survey and describe the properties and spatial distributions of these new clusters in Section 4. In Section 5, we discuss the spatial distribution of the new clusters in comparison with previously known clusters in the survey region. Our primary results are summarized in Section 6.

\section{Data}
\begin{figure*}
\epsscale{1.2} 
\plotone{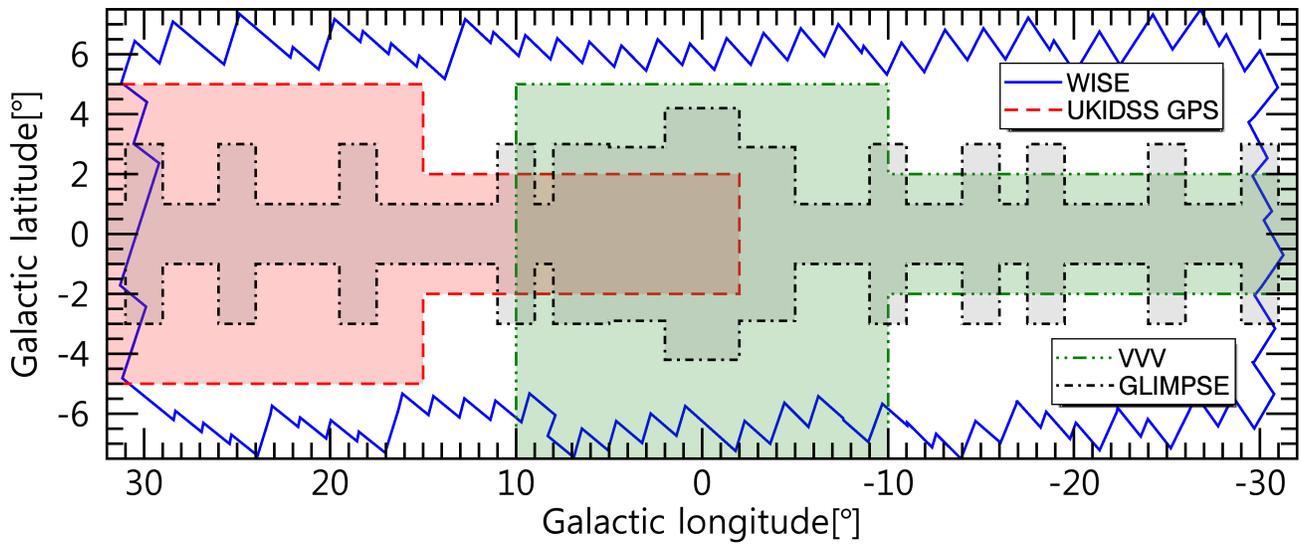}
\caption{Schematic area coverages of wide-field surveys. The jagged solid line represents the area of the WISE survey used in this study. Dashed, dash-triple dot, and dash-dot lines represent the UKIDSS GPS, VVV, and GLIMPSE survey area, respectively. \label{schematic}}
\end{figure*}

Our main survey data are based on the WISE. The WISE survey covers the entire sky with four filters, $3.3\mu\mathrm{m}$ (W1), $4.6\mu\mathrm{m}$ (W2), $12\mu\mathrm{m}$ (W3), and $22\mu\mathrm{m}$ (W4). We select a region with $|l|<30\arcdeg$ and $|b|<6\arcdeg$ centered on the Galactic center for our WISE cluster survey, covering a much larger area than previous infrared cluster surveys. We obtain a section image with 90$\arcmin\times$90$\arcmin$ field of view from the WISE survey and retrieve 334 such section images for each filter. We display a schematic area coverage of our survey in Figure \ref{schematic}, in comparison with other surveys (UKIDSS Galactic Plane Survey (GPS; \citealt{luc08}), VVV, and GLIMPSE).

The WISE data suffer less extinction problems than the 2MASS. Because of even the W1 band, which has the shortest central wavelength among the four bands of the WISE, has a longer central wavelength than the $K_s$ band. However, it is hard to derive the properties of clusters using only the WISE data. The spatial resolving power of the WISE is weaker than that of the 2MASS. The FWHM of point sources is as large as $6\arcsec.1$ in the W1 band, which is much larger than that of the 2MASS $K_s$ band, $2\arcsec.6$. The numbers of point sources detected in images of the W3 and W4 bands are not large enough to derive the photometric properties of clusters. Moreover, the intrinsic (W1--W2) colors of typical stars are almost constant around --0.1. Therefore, other NIR survey data to support the WISE data are needed.

The 2MASS provides useful NIR information through three bands ($JHK_s$). However, the limiting magnitude of the 2MASS for our survey region is relatively bright, $K_s\simeq15$ mag. This limit is not faint enough to investigate clusters located deep inside the Galactic plane.

Thus, we also use two NIR surveys deeper than the 2MASS focusing on the Galactic plane: the UKIDSS GPS and the VVV. These two surveys were conducted with 4-m telescopes, so the limiting magnitudes of those surveys are $\sim3$ mag deeper than those of the 2MASS. The UKIDSS GPS covers the first and second quadrant of the Galactic plane: ($|b|<2\arcdeg$ and $-2\arcdeg<l<15\arcdeg$) and ($|b|<5\arcdeg$ and $15\arcdeg<l<107\arcdeg$). On the other hand, the VVV covers the fourth quadrant of the Galactic plane: ($-10\arcdeg<b<5\arcdeg$ and $-10\arcdeg<l<10\arcdeg$) and ($|b|<2\arcdeg$ and $-65\arcdeg<l<-10\arcdeg$). These surveys overlap about two-thirds of our survey region.

Additionally, we also use the GLIMPSE data to supplement the WISE data. The effective wavelength of the W1 filter of the WISE is similar to that of the [3.6] filter of the \textit{Spitzer} telescope. However, the FWHM of the W1 ($6\arcsec.1$) is much larger than that of the \textit{Spitzer}, $2\arcsec.0$ in the [3.6] band. Hence the GLIMPSE data are more useful than the WISE data for counting the number of sources in the similar wavelength range. However, the GLIMPSE data cover only the $|b|<2\arcdeg$ region.

\section{Method}
The process of finding star clusters in this study consists of two steps. First, we visually search for cluster candidates in the entire survey region using the WISE W1 images. Second, we classify these candidates based on three criteria: the morphological features seen in the W1 image, the Color-Magnitude Diagrams (CMDs) of stars based on the NIR data, and the Radial number Density Profiles (RDPs) of stars derived from the $K_s$ and the W1 band (or the [3.6] band of the GLIMPSE). Details for these procedures are described in the following.

\subsection{Visual Search for Cluster Candidates}
Most of the cluster findings in previous studies focused on the overdense regions found from the catalogs of point sources. Several surveys found overdense regions and removed spurious candidates among them by visual checking \citep[e.g.][]{mer05, fro07, bor11, sol12, mor13, bor14, sol14, bar15}. However, this cluster finding method is not much useful in the case of the WISE, because WISE photometry is not deep enough. Therefore, we investigate the morphological features of extended sources including some point sources in the WISE W1 images visually. We use three criteria to find cluster candidates: (1) the apparent size of the candidates, (2) the surface brightness of the candidates, and (3) the apparent angular separation from known clusters in the literature.

First, we choose stellar groups with a radius smaller than 5$\arcmin$ as cluster candidates. This size limit is a guideline for our cluster candidate finding. Note that the angular radii of all known globular clusters, known embedded clusters, and 72\% of known open clusters appear to be smaller than 5$\arcmin$ in the sky \citep{har96, lad03, dia02}, regardless of their distances. 

Second, we use W1 surface brightness maps to check the cluster candidates. These surface brightness maps are made with W1 images that were smoothed using the Gaussian kernel with 
$180\arcsec$ FWHM. This FWHM size for the Gaussian kernel is set to reveal better stellar groups with a few arcmin radius.
In fact, known globular clusters and embedded clusters are easily distinguishable in these surface brightness maps.
In addition, we focus on the dips in the surface brightness maps. There are some cases that only a few stars embedded in diffuse light are seen around the surface brightness dips. These dips represent severely obscured regions, which may host embedded clusters.

Third, we select cluster candidates that are not overlapped with the known clusters. References of the known clusters are listed in Table \ref{t1}. In our survey region, there are 860 known clusters in total, including 38 globular clusters \citep{har96}, 244 open clusters \citep{dia02}, and 252 embedded clusters (i.e. \citealt{dut03, bic03b, mer05, bor11, mor13}).

This cluster candidate selection process is likely to be biased in selecting the groups of NIR-bright stars because of the characteristics of the WISE data. These NIR-bright stars are likely to be red giants, red supergiants, or pre-main sequence stars rather than the NIR-faint main sequence stars. As a consequence, the cluster candidates are likely to be old clusters with well-developed red giants, young star clusters with red supergiants, or young embedded clusters showing the infrared excess that comes from pre-main sequence stars.

\subsection{Classification of the Cluster Candidates}
\begin{figure*}
\plotone{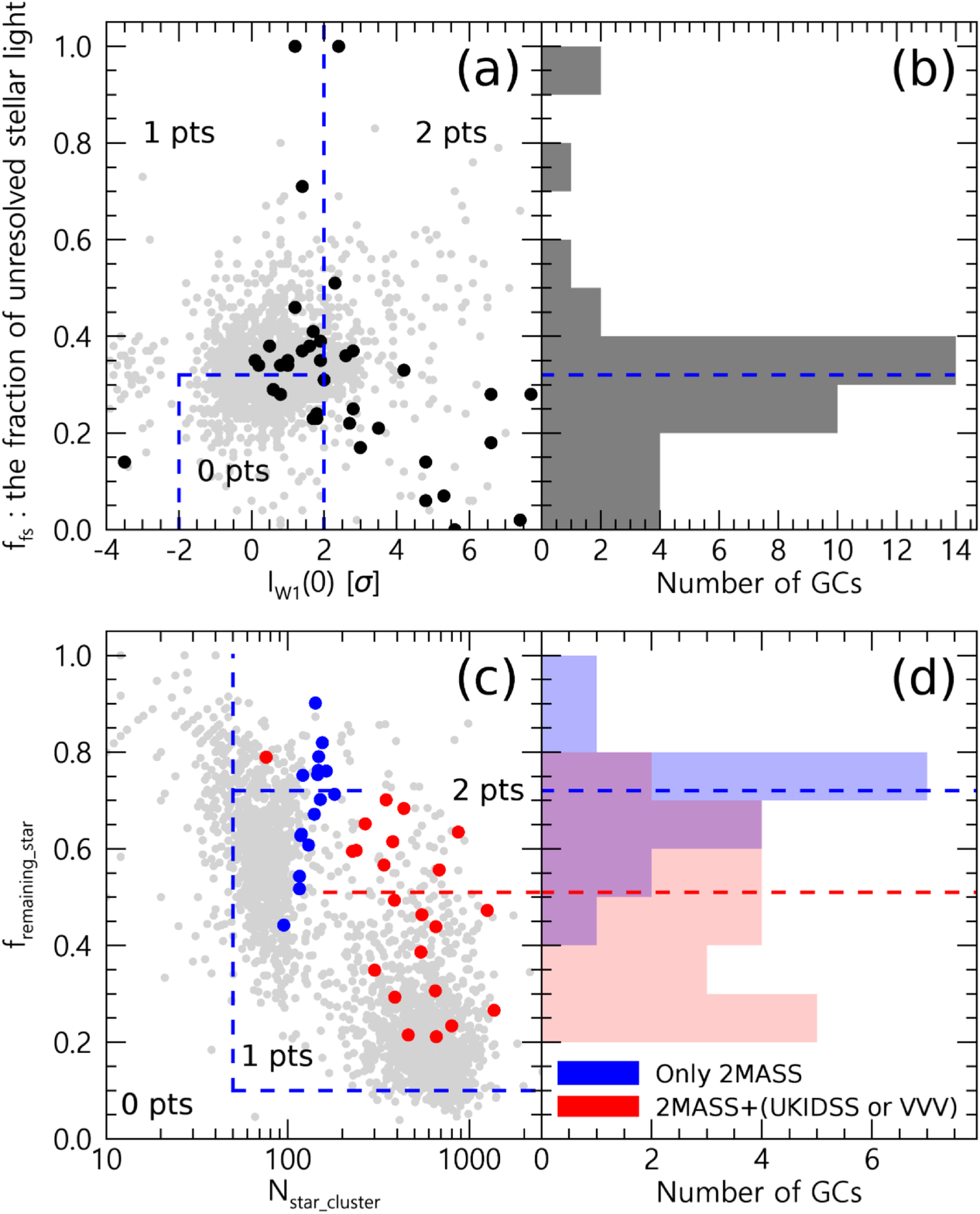}
\caption{(a) A morphological classification diagram 
for 38 globular clusters (GCs; filled circle) in the Milky Way. Gray points are the cluster candidates found in this study.
(b) A number distribution of $f_{fs}$ values for the GCs. 
The dashed line represents a median value, 0.32. (c) A CMD classification diagram for the same GCs.
(d) A number distribution of $f_{remaning\_star}$ values for the GCs. We divide GCs into two subgroups: with deep NIR data (red filled circle) and without deep NIR data (blue filled circle). Each subgroup has different median $f_{remaning\_star}$ values, 0.72 for the GCs without deep NIR data and 0.51 for the GCs with deep NIR data. Dashed lines represent median values for the each subgroup. \label{classify}}
\end{figure*}

After selecting the cluster candidates, we grade them according to three criteria: morphology, CMDs, and RDPs of the candidates. These criteria and the grading system are described in the following, are shown in Figure \ref{classify}, and are summarized in Table \ref{t2}.

\begin{figure*}
\plotone{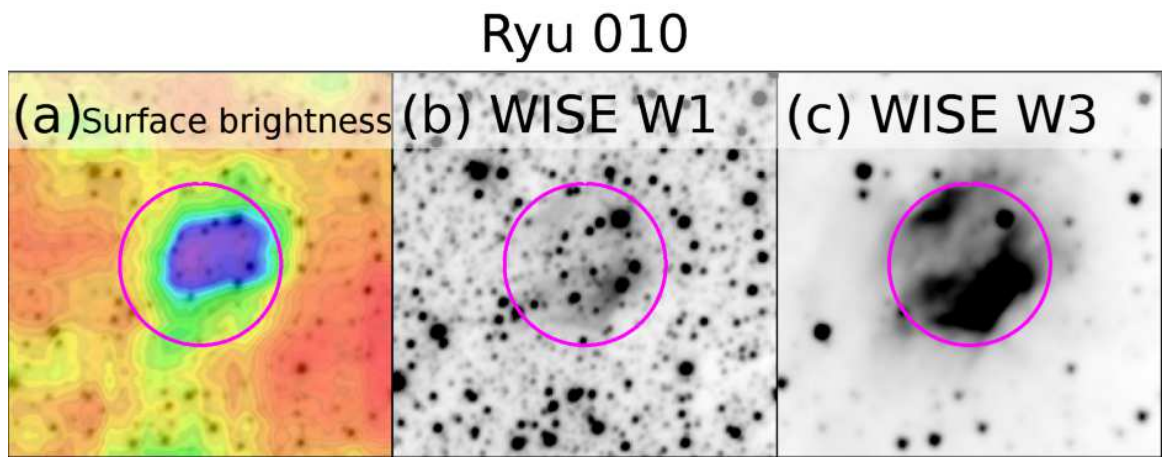}
\caption{(a) Smoothed surface brightness map of W1 image, (b) W1 image, and (c) W3 image of a new cluster sample, Ryu 010. The field of view of each image is $10\arcmin\times10\arcmin$. Circles represent the apparent size of this cluster ($R=2\arcmin.3$). North is up, and east is left. \label{ryu010img}}
\end{figure*}

\textit{(i) Morphology.} We define two parameters for the quantitative morphology grading that 
are based on the W1 image. One is the central surface brightness ($I_{W1}(0)$; Figure \ref{ryu010img}(a)) relative to the background value $I_{W1,bg}$. 
The other is the faint star ratio that is defined as $f_{fs}=N_{pixel}(r<R_{cl}, I_{W1,bg}\leq I_{W1}\leq I_{W1,bg}+2\sigma_{bg}) / N_{pixel}(r<R_{cl})$, where $I_{W1}$ and $I_{W1,bg}$ are the value of a pixel and the median value of the background region ($3R_{cl}<r<3.6R_{cl}$). This $f_{fs}$ 
represents the spatial coverage of the unresolved stellar light in the cluster region.

The morphological classification using these two parameters is described in Figure \ref{classify}(a). We adopt $f_{fs}=0.32$ as a reference point for grading because it is the median value of $f_{fs}$ for the globular clusters in our survey area, as shown in Figure \ref{classify}(b).

In addition, we classify embedded clusters using the existence of diffuse light seen in the W1 and W3 images of the cluster candidates (Figure \ref{ryu010img}(b) and (c)). We checked visually any feature of diffuse light around the target object notable in the W1 and W3 images at the same time. These diffuse light in the W1 and W3 images might be originated from the Polycyclic Aromatic Hydrocarbon (PAH) emission features, which is known to be a dust tracer \citep{wri10}. According to the definition of embedded clusters \citep{lad03}, the existence of the IR diffuse light is a good indicator of embedded clusters. This embedded classification is independent of the quantitative classification.

\begin{figure}
\epsscale{1.15} 
\plotone{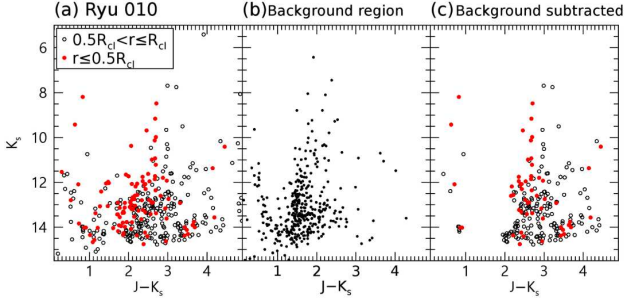}
\caption{(a) The $K_s$--$(J-K_s)$ CMD of Ryu 010 ($R=2\arcmin.3$). Open and filled circles represent stars located at $0.5R_{cl}<r\leq R_{cl}$ and $r\leq0.5R_{cl}$, respectively. (b) The $K_s$--$(J-K_s)$ CMD of the background region of Ryu 010 ($3R_{cl}<r<3.6R_{cl}$). The area of this background region is 4 times wider than the cluster region. Only a quarter of the stars are randomly selected in the background region for plotting the CMDs. (c) The background-subtracted $K_s$--$(J-K_s)$ CMD. The symbols are same as in (a). \label{ryu010cmd}}
\end{figure}

\textit{(ii) CMDs.} We use the $K_s$--$(J-K_s)$ CMD of stars in the cluster candidate region and the background region for grading. The background region is an annular region whose area is 4 times larger than that of the cluster candidate region. We set the inner and outer radii of the background region as 3 and 3.6 times the radii of the cluster candidates, respectively. As an example, the CMDs of Ryu 010 are shown in Figure \ref{ryu010cmd}. 
We use 2MASS photometry for CMDs. When the UKIDSS GPS or the VVV data are available, we combine them with the 2MASS data. For this case, we use the 2MASS data for stars brighter than $K_s=12$ mag, and the UKIDSS GPS or the VVV data for fainter stars. We use the two following features in the CMD for grading: the number of stars of the cluster candidate region ($N_{star\_cluster}$; Figure \ref{ryu010cmd}(a)) and the ratio of remaining stars after the statistical background subtraction process ($f_{remaining\_star}$; Figure \ref{ryu010cmd}(c)) that is based on the number difference of stars between the cluster candidate region and the background region (Figure \ref{ryu010cmd}(b)) in the CMDs. The statistical background subtraction process is described in Appendix B.

The CMD classification using those two parameters is described in Figure \ref{classify}(c). The distribution of $f_{remaining\_star}$ depends on the existence of deep NIR data, because faint stars are likely to be removed by the statistical subtraction process. Therefore we adopt two different reference points for grading cluster candidates. The reference points, median values of $f_{remaining\_star}$ for the globular clusters in our survey area are 0.72 (without deep NIR data) and 0.51 (with deep NIR data), as shown in Figure \ref{classify}(d).

\begin{figure}
\epsscale{0.9}
\plotone{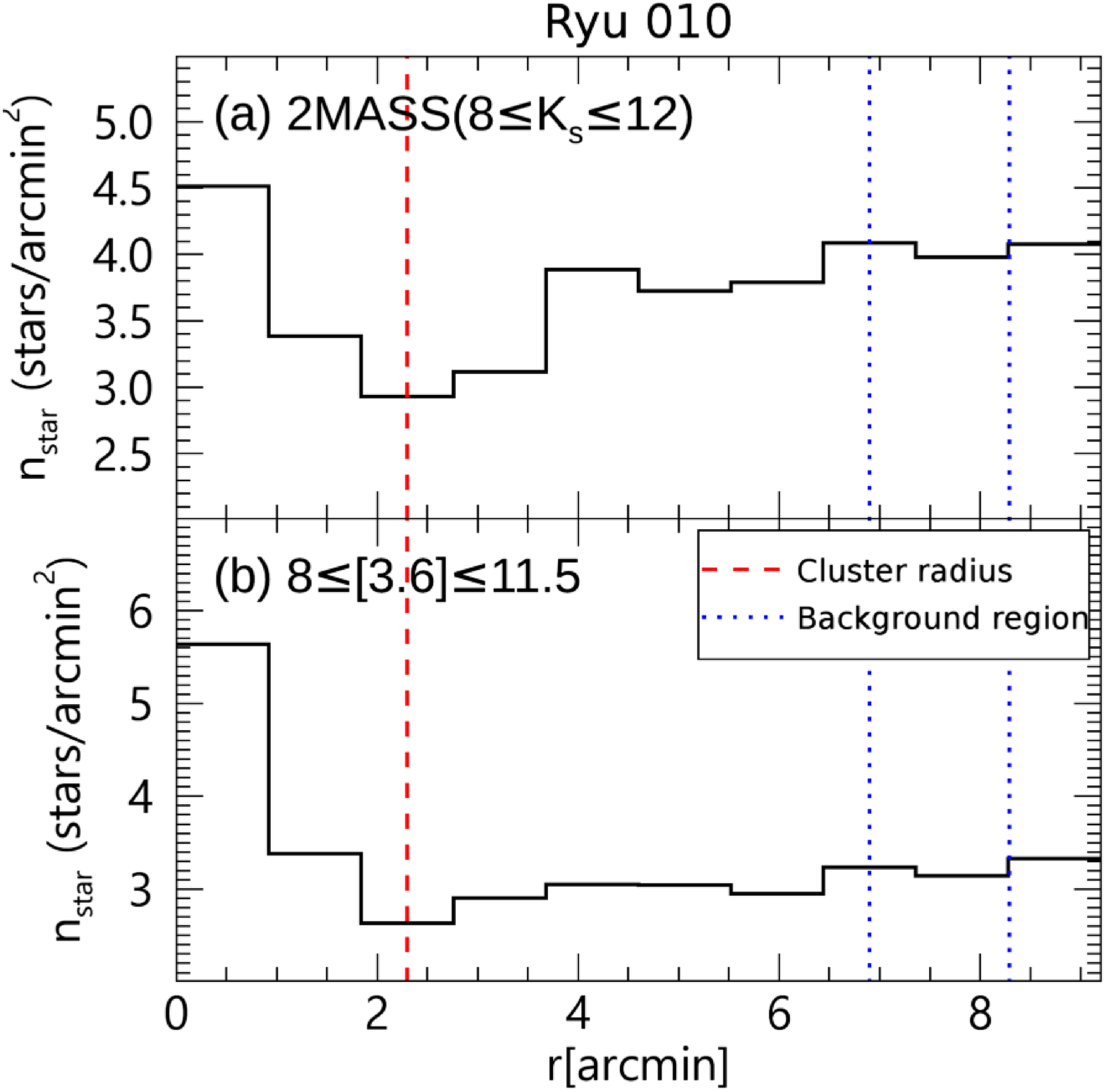}
\caption{(a) An RDP of Ryu 010 derived from the $K_s$ band data. The 2MASS data is used for stars with $8\leq$ $K_s\leq12$ mag. The dashed line represents the radius of this cluster ($R=2\arcmin.3$). Two dotted lines represent the inner and outer radii of the background region ($3R_{cl}<r<3.6R_{cl}$) of Ryu 010. (b) An RDP derived from the GLIMPSE [3.6] band data. Stars with $8\leq$ $[3.6]\leq11.5$ mag are used. \label{ryu010rdp}}
\end{figure}

\textit{(iii) RDPs.} It is expected that clusters show a number density excess at the center of the cluster. We derive RDPs in two bands to check the central number density excess: one based on $K_s$ band data (2MASS data, UKIDSS GPS, or VVV data), and the other based on the W1 or GLIMPSE [3.6] bands. The central excess is defined as $n_{peak} > n_{bg}+2\sigma_{star\_bg}$, where the $n_{peak}$, $n_{bg}$, and $\sigma_{star\_bg}$ are the number density peak in the cluster region, the mean number density of the background region, and the standard deviation of the number density of the background region, respectively. Figure \ref{ryu010rdp} shows an example of RDPs of Ryu 010. In this figure, the central excess is seen in both RDPs. 

The radii of clusters are determined using the RDP at the W1 or the [3.6] band, which is the point where the number density drops to the background level. We adopt $\pm0.2R_{cl}$ as the radius error for all clusters in common, corresponding to the width of each bin in the RDP. 

We grade not only the cluster candidates but also the known clusters in our cluster survey field. For each of the three criteria, the targets are given 0, 1, or 2 points. Then the total points of the cluster candidates are derived by summing up the points. We consider the graded targets with 4 or more total points as clusters and divide these clusters into three classes: A, B, and C for the clusters with 6, 5, and 4 points, respectively. In addition, the embedded clusters with points lower than 4 points are included in class C.

To check the validity of our grading system, we investigate the known clusters covered in our survey regions; there are 38 globular clusters, 244 open clusters, and 252 embedded clusters. The average points of the globular clusters and the open clusters are $4.68\pm0.93$ and $3.30\pm1.24$, respectively. The globular clusters show the highest average point among the known clusters. Almost all globular clusters get 4 or more total points, which means that $89\%$ (34/38) of the globular clusters are recovered by this study. In contrast, the average point of the open clusters means that only $48\%$ of them are recovered. It is because the open clusters are not bright in the NIR regime during most of their lifetime. However, we cannot compare the known embedded clusters with the other clusters by average points due to the different classification criteria. The average point of the known embedded clusters is $3.79\pm1.49$, which is consistent with the value of the open clusters and much smaller than the value of the globular clusters. Nevertheless, it is noted that we classify the embedded clusters by the diffuse light in the cluster region, not by the total points. According to these results, we conclude that our grading system is effective for classifying clusters: old clusters by the total points and embedded clusters by the presence of IR diffuse light.

We also check our grading system using the False Positive Rate (FPR). FPR is derived using 1000 `false samples', which are not clusters but randomly selected field regions. The false sample follows the spatial and size distributions of the new clusters. However, they do not overlap with the new 
or known clusters. We grade the false sample 
in the same way as 
for the new clusters. The average point of the false sample is $1.95\pm1.05$. 
This value is much lower 
than the cluster classification criterion. FPR, the ratio of the false sample which get 4 or higher scores to the entire set of the false sample, is $6.4\%$. 
This shows that the effect of contamination in our new cluster sample is minor.

\section{Results}
\subsection{A Catalog of New Clusters}
From the 1722 cluster candidates, we finally select 923 clusters and list them in Table \ref{t3}. They consist of 15 class A, 254 class B, and 654 class C clusters. We find that 202 ($22\%$) of these 923 clusters are classified as embedded clusters. The number of embedded clusters in each class is 5 ($33\%$; 5/15), 35 ($14\%$; 35/254), and 162 ($25\%$; 162/654) for the classes A, B, and C, respectively.

We cross-match these new clusters with various sources in the literature: 412 ($45\%$) the Infrared Astronomical Satellite (IRAS) point sources \citep{bei88}, 767 ($83\%$) the Midcourse Space Experiment (MSX) sources \citep{ega03}, 507 ($55\%$) dark nebulae \citep{per09, dob11, dut02}, 201 ($22\%$) young stellar objects \citep{tot14, rob08, mot07}, 30 ($3\%$) methanol masers \citep{pes05, cas10, gre10, cas11, sun14}, 32 ($3\%$) water masers \citep{wal14, sun07, urq11}, 14 ($2\%$) OH masers \citep{sev97a, sev97b, sev01, tel91}, and 10 ($1\%$) SiO masers \citep{har98, yoo14, nak07}. In Table \ref{t4}, we list cross-matching results of the new clusters.

A number of clusters are cross-matched with the IRAS point sources, MSX sources, and dark nebulae. The cross-matching result for these sources is summarized in Table \ref{t5}. In the case of the IRAS point sources, 10 ($67\%$; 10/15), 108 ($43\%$; 108/254), and 294 ($45\%$; 294/654) clusters are cross-matched in class A, B, and C, respectively. Most of these IRAS point sources are cross-matched with the embedded clusters; the number of cross-matched sources within the embedded clusters are 4 ($80\%$; 4/5) for class A, 26 ($74\%$; 26/35) for class B, and 126 ($78\%$; 126/162) for class C. The MSX sources show a stronger concentration with the new clusters than the IRAS point sources; 14 ($93\%$; 14/15), 216 ($85\%$; 216/254), and 537 ($82\%$; 537/654) clusters are cross-matched in class A, B, and C, respectively. Moreover, the MSX sources are almost same as the embedded clusters; 5 ($100\%$; 5/5) for class A, 34 ($97\%$; 34/35) for class B, and 157 ($97\%$; 157/162) for class C. In consequence, for dark nebulae, 10 ($67\%$; 10/15), 129 ($51\%$; 129/254), and 368 ($56\%$; 368/654) clusters are cross-matched in class A, B, and C, respectively. Also, those nebulae are coincident with the embedded clusters; 4 ($80\%$; 4/5) for class A, 30 ($86\%$; 30/35) for class B, and 133 ($82\%$; 133/162) for class C.

\subsection{Sizes of the New Clusters}
\begin{figure}
\plotone{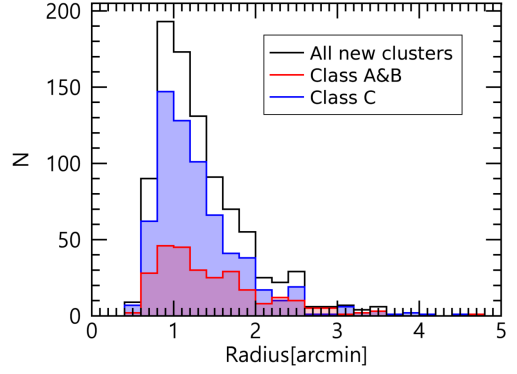}
\caption{The size distribution of the new clusters. The empty black histogram represents the entire set of new clusters. The red-shaded histogram represents the class A and B clusters, and the blue-shaded histogram represents the class C clusters. \label{radhist}}
\end{figure}

Although we select cluster candidates with radii smaller than 5$\arcmin$, finally chosen clusters tend to be more compact than this value. The size distribution of the new clusters is shown in Figure \ref{radhist}. As can be seen, 902 ($98\%$) clusters have radii smaller than $R=3\arcmin$, and the radii of 823 ($89\%$) clusters are even smaller than $R=2\arcmin$. The average radius of the clusters is $1\arcmin.31\pm0\arcmin.60$. In the survey region, small and compact clusters are easier to find visually than large and sparse clusters, because the former suffer from the background star contamination less than the latter. In contrast, the radii of 921 ($99.8\%$) clusters are larger than $R=0\arcmin.5$. We consider this value of $R=0\arcmin.5$ is a lower size limit of the cluster classification, because this value is only 5 times of the FWHM size of the W1 images. The clusters smaller than $R=0\arcmin.5$ are difficult to classify.


\subsection{Spatial Distributions of the New Clusters}
\begin{figure*}
\epsscale{1.2}
\plotone{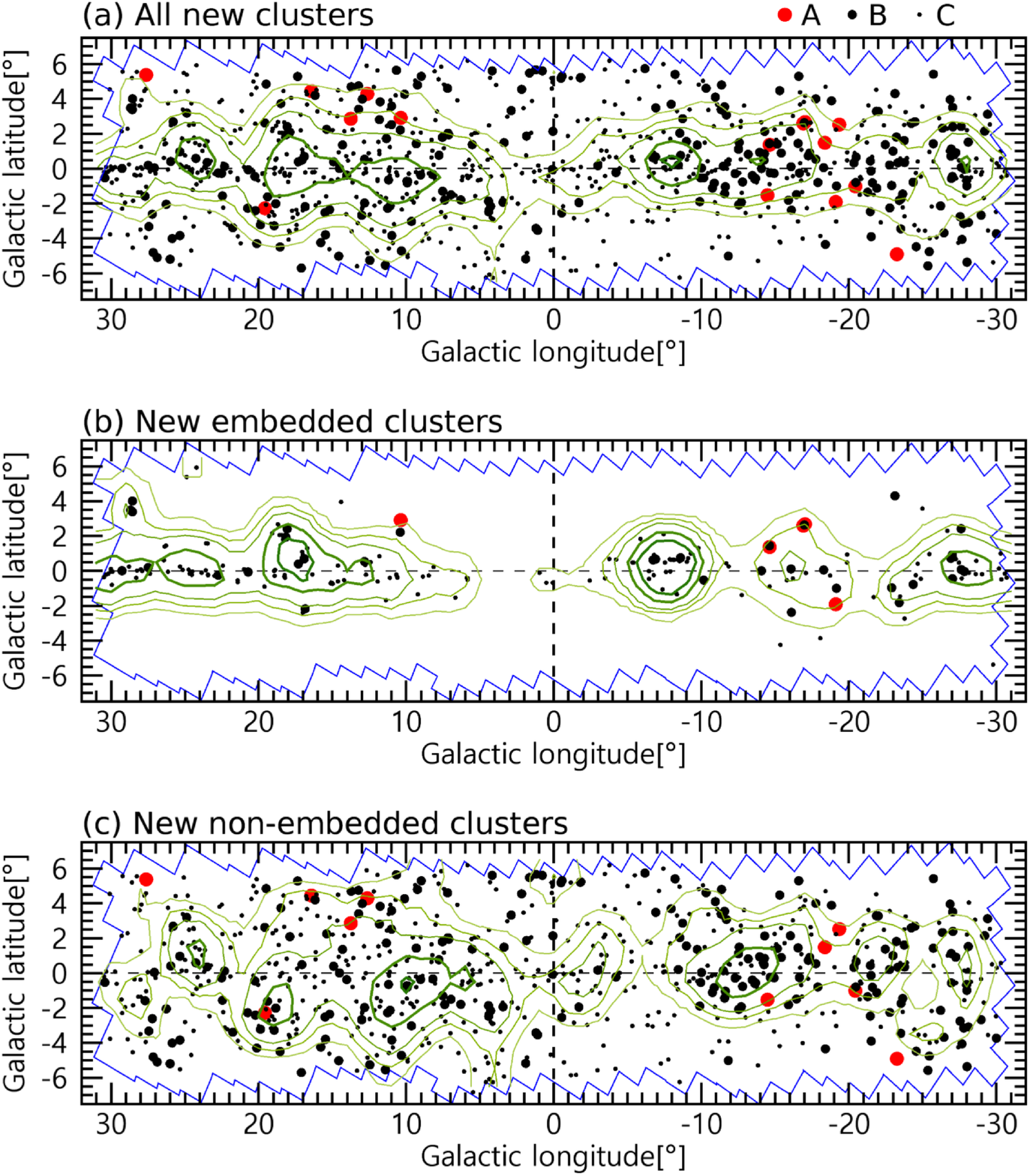}
\caption{Spatial distributions of (a) the new clusters, (b) the embedded clusters, and (c) the non-embedded clusters. The size of symbols represents their classes: class A clusters are the biggest and class C clusters are the smallest. Contours represent the number density of the new clusters. Contour levels represent 0$\sigma$, 0.5$\sigma$, 1$\sigma$, 2$\sigma$, and 3$\sigma$ excess with respect to the average number density of the entire survey region. Thick contours are used to emphasize 2$\sigma$ and 3$\sigma$ excess. \label{spd-cands}}
\end{figure*}

\begin{figure*}
\epsscale{1.2}
\plotone{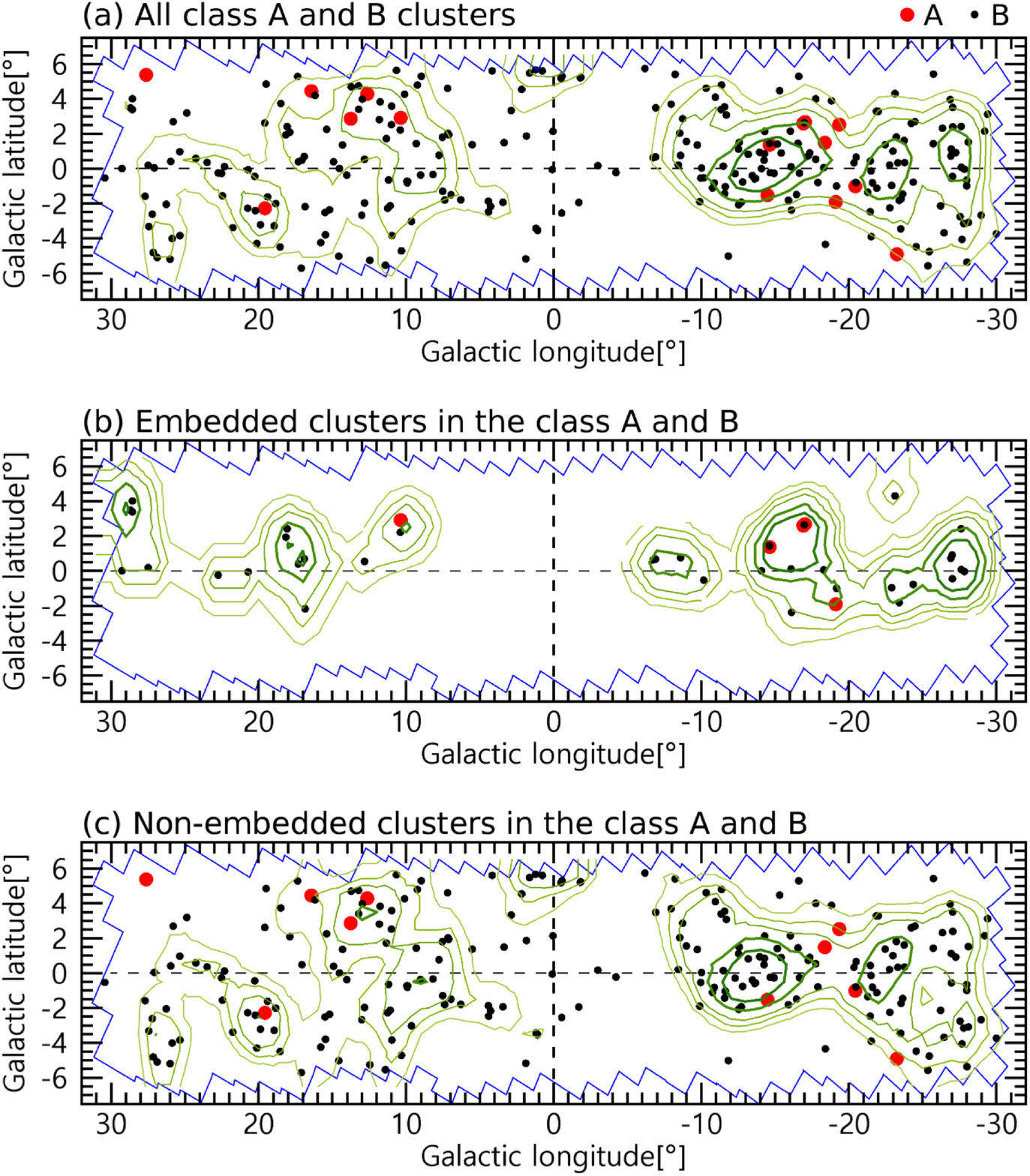}
\caption{Spatial distributions of (a) all class A and B clusters, (b) embedded clusters in the class A and B, and (c) non-embedded clusters in the class A and B. Red and black filled circles represent the class A and B clusters, respectively. Contours and lines are same as in Figure \ref{spd-cands}. \label{spd-ab}}
\end{figure*}

\begin{figure*}
\epsscale{1.2}
\plotone{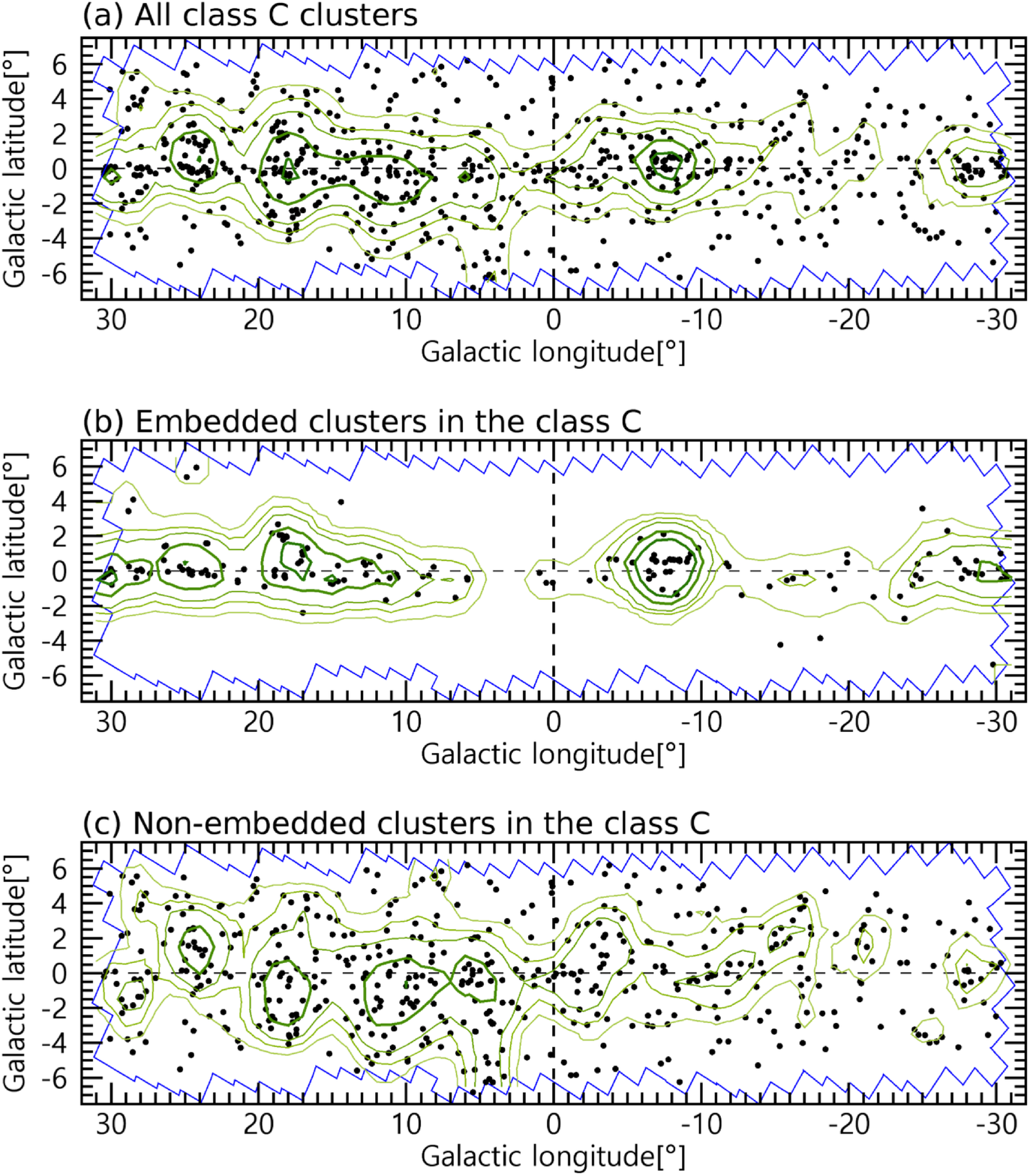}
\caption{Spatial distributions of (a) all class C clusters, (b) embedded clusters in the class C, and (c) non-embedded clusters in the class C. Contours and lines are same as in Figure \ref{spd-cands}. \label{spd-c}}
\end{figure*}

Figure \ref{spd-cands} shows spatial distributions of the entire new clusters in the sky, and Figure \ref{spd-ab} and \ref{spd-c} display spatial distributions of the class A+B and class C clusters, respectively. Contour maps of the number density of the entire new clusters (Figure \ref{spd-cands}(a)) show a clumpy structure. We divide the entire set of new clusters into two groups, embedded and non-embedded clusters, and investigate their spatial distributions separately. In Figure \ref{spd-cands}(b), the embedded clusters show strong concentration around the Galactic plane. They are not continuously distributed along the Galactic plane. There are only a few embedded clusters around the $-3\arcdeg<l<8\arcdeg$ region. However, some peaks of the number density are seen at the $(l, b) \simeq $ ($-28\arcdeg$, $0\arcdeg$), ($-8\arcdeg$, $0\arcdeg.5$), ($17\arcdeg.5$, $1\arcdeg$), ($24\arcdeg$, $0\arcdeg$), and ($29\arcdeg$, $0\arcdeg$). In contrast, non-embedded clusters, in Figure \ref{spd-cands}(c), show weaker concentrations toward the Galactic plane. All peaks in Figure \ref{spd-cands}(b) are not visible but other peaks are revealed at the $(l, b)\simeq $ ($-13\arcdeg$, $0\arcdeg$), ($10\arcdeg$, $0\arcdeg.5$), ($19\arcdeg$, $-2\arcdeg$), and ($24\arcdeg$, $1\arcdeg.5$).

The differences between the embedded and non-embedded clusters are also visible in the spatial distribution of the class A+B clusters. Among the several peaks shown in Figure \ref{spd-cands}(b), the $(l, b) \simeq $ ($-28\arcdeg$, $0\arcdeg$), ($-8\arcdeg$, $0\arcdeg.5$), and ($17\arcdeg.5$, $1\arcdeg$) peaks appear in the embedded clusters of class A+B, in Figure \ref{spd-ab}(b). Despite small numbers, two more peaks are revealed at the $(l, b) \simeq$ ($-17\arcdeg$, $1\arcdeg$) and ($29\arcdeg$, $4\arcdeg$) that are not obvious in Figure \ref{spd-cands}(b). The last peak is located close to the position of the Aquila Rift, a famous molecular cloud complex. On the other hand, the non-embedded clusters in class A+B are distributed more broadly than the embedded clusters in the same class, as shown in Figure \ref{spd-ab}(c). Only the strongest peak at the $(l, b)\simeq $ ($-13\arcdeg$, $0\arcdeg$) in Figure \ref{spd-ab}(c) is shown in common with Figure \ref{spd-cands}(c). There are two more weak peaks 
 in the spatial distribution of the non-embedded clusters in class A+B: at the $(l, b) \simeq $ ($-23\arcdeg$, $1\arcdeg$) and ($13\arcdeg$, $3\arcdeg.5$).

The distributions of the class C clusters shown in Figure \ref{spd-c} are similar to the distributions of the entire set of new clusters, new embedded clusters, and new non-embedded clusters as shown in Figure \ref{spd-cands}. It is because a majority ($71\%$) of the new clusters belong to class C. However, the distributions of the class C clusters are different from the class A+B clusters. 
In the case of the embedded clusters in class A+B and class C, a prominent peak at the $(l, b) \simeq$ ($-8\arcdeg$, $0\arcdeg.5$) is coincident in Figure \ref{spd-ab}(b) and Figure \ref{spd-c}(b). A broad number excess region at the $23\arcdeg<l<30\arcdeg$ shown in the spatial distributions of class C becomes negligible in class A+B. On the other hand, no spatial peaks of the non-embedded clusters in class C at the $(l, b) \simeq $ ($5\arcdeg$, $0\arcdeg$), ($10\arcdeg$, $-1\arcdeg$), ($18\arcdeg$, $-1\arcdeg$), and ($24\arcdeg$, $1\arcdeg.5$) are coincident with the non-embedded clusters in class A+B.

\subsection{Case studies of 15 class A clusters using CMDs}
\begin{figure*}
\epsscale{1.1} 
\plotone{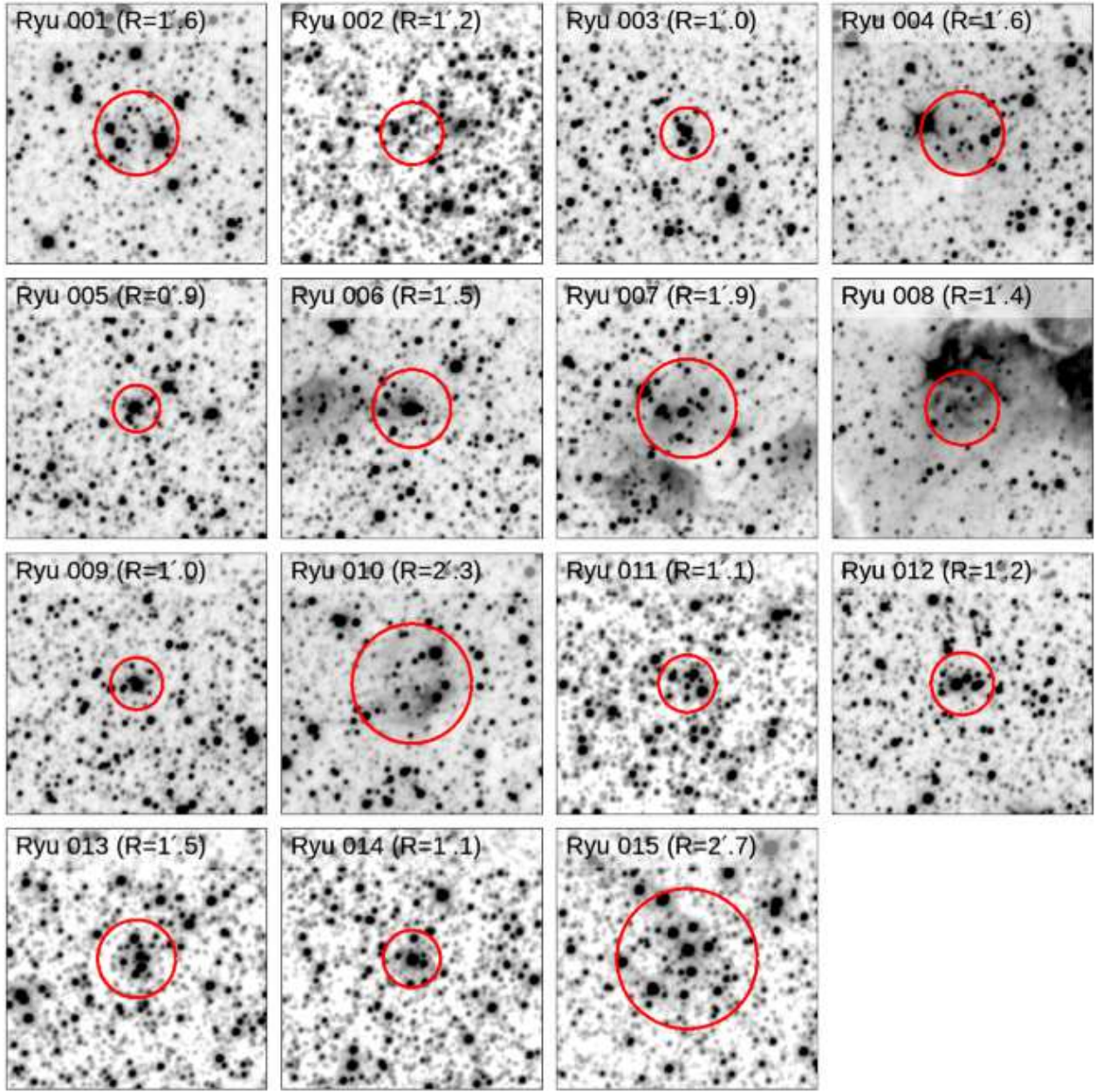}
\caption{Gray-scale maps of W1 band images of 15 class A clusters. The field of view of each image is $10\arcmin\times10\arcmin$. Circles represent the apparent sizes of the clusters. The radius of the cluster is given in each map. North is up, and east is left. \label{cand_sel}}
\end{figure*}

We estimate reddenings, distances, and relative ages of 15 class A clusters, 
of which gray-scale W1 band images are shown in Figure \ref{cand_sel}. These physical parameters are determined using the PARSEC isochrone \citep{bre12, mar08}, with assumptions of the solar metallicity and the total-to-selective extinction ratio $R_V =3.1$. We fit visually the cluster sequence with the isochrones on the statistically background-subtracted CMDs. First, the reddening of a cluster is determined on the $K_s$--$(H-K_s)$ CMD. In this CMD, the stellar locus is vertically stretched regardless of the distance and age. Second, the distance to a cluster is determined on the $K_s$--$(J-K_s)$ CMD. In the $K_s$--$(J-K_s)$ CMD, red giant branches (RGB) of old stars or the pre-main sequences are tilted. Therefore, we can estimate the cluster distances from the magnitudes of the tip of the RGB (TRGB) stars for old clusters or from the magnitudes of the pre-main sequence stars for embedded clusters. However, it should be kept in mind that these distances are highly uncertain. For old clusters, it is difficult to determine the precise position of the TRGB because there are only a small number of TRGB stars in an individual cluster. For embedded clusters, the pre-main sequence stars are a less reliable distance indicator than the main sequence stars, because the colors and magnitudes of the pre-main sequence stars vary through the ages. 

Only relative age estimation is possible for these clusters. In the case of embedded clusters, we need to consider the following effects to determine their ages from isochrone fitting: the internal age difference, the differential reddening, the distance uncertainty, and the inaccuracy of the stellar evolution model at the pre-main sequence stage. Because of these reasons, we use 4 Myr isochrones as a reference for the age of the embedded clusters. In the case of old clusters, their turn-off points are too faint to determine ages. Nevertheless, they seem to be older than 800 Myr, because the RGB-like sequence is seen in their CMDs. Therefore we use 10 Gyr isochrones for old clusters generally, except for a few clusters that show subgiant branch stars marginally. For those clusters, we use the 800 Myr and 2 Gyr isochrones. As a result, we divide the clusters into 5 embedded clusters and 10 old clusters.

\begin{figure*}
\epsscale{1.1} 
\plotone{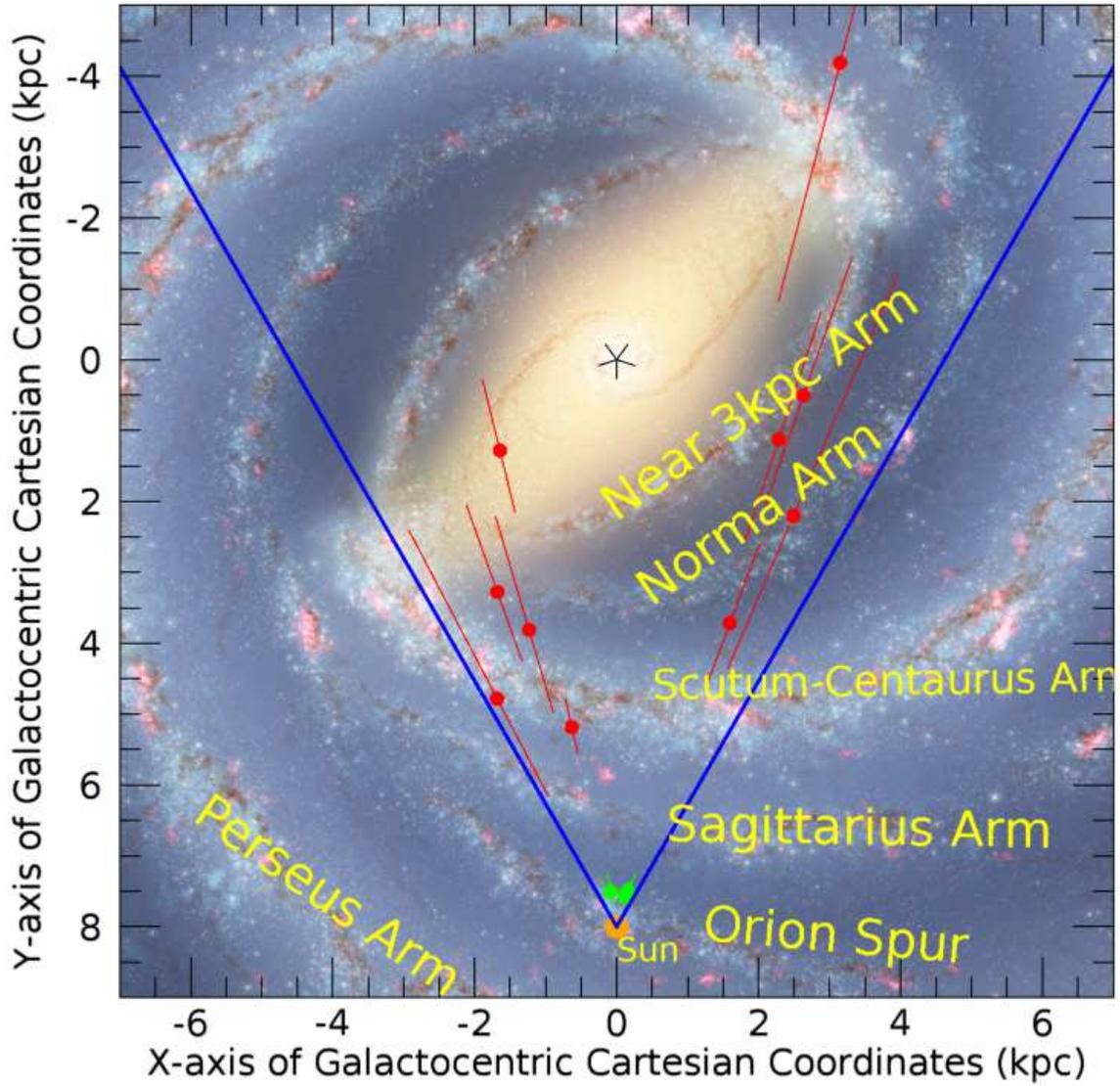}
\caption{Spatial distribution of 15 class A clusters on the face-on view of the Milky Way \citep{chu09}. The green and red points with error bars represent the locations of the young and old clusters, respectively. Two blue solid lines represent the spatial boundary of our cluster survey region. \label{spd-mwp}}
\end{figure*}

The estimated physical parameters are listed in Table \ref{t6}. $E(B-V)$ values range from 0.2 to 5.2, and distances range from 440 pc to 12.6 kpc. We estimate the spatial distribution of the 15 class A clusters on the face-on view of the Milky Way using positions and distances of these clusters as shown in Figure \ref{spd-mwp}. The embedded clusters are located closer than the old clusters. 
Due to the faint absolute magnitudes of the pre-main sequence stars, it is 
difficult to discover distant embedded clusters.
In contrast, 
RGB stars in old clusters are much brighter than the pre-main sequence stars, so we can find old clusters at much larger distances than the embedded clusters.

\section{Discussion}
\subsection{Comparison with known clusters}
\begin{figure*}
\epsscale{1.1} 
\plotone{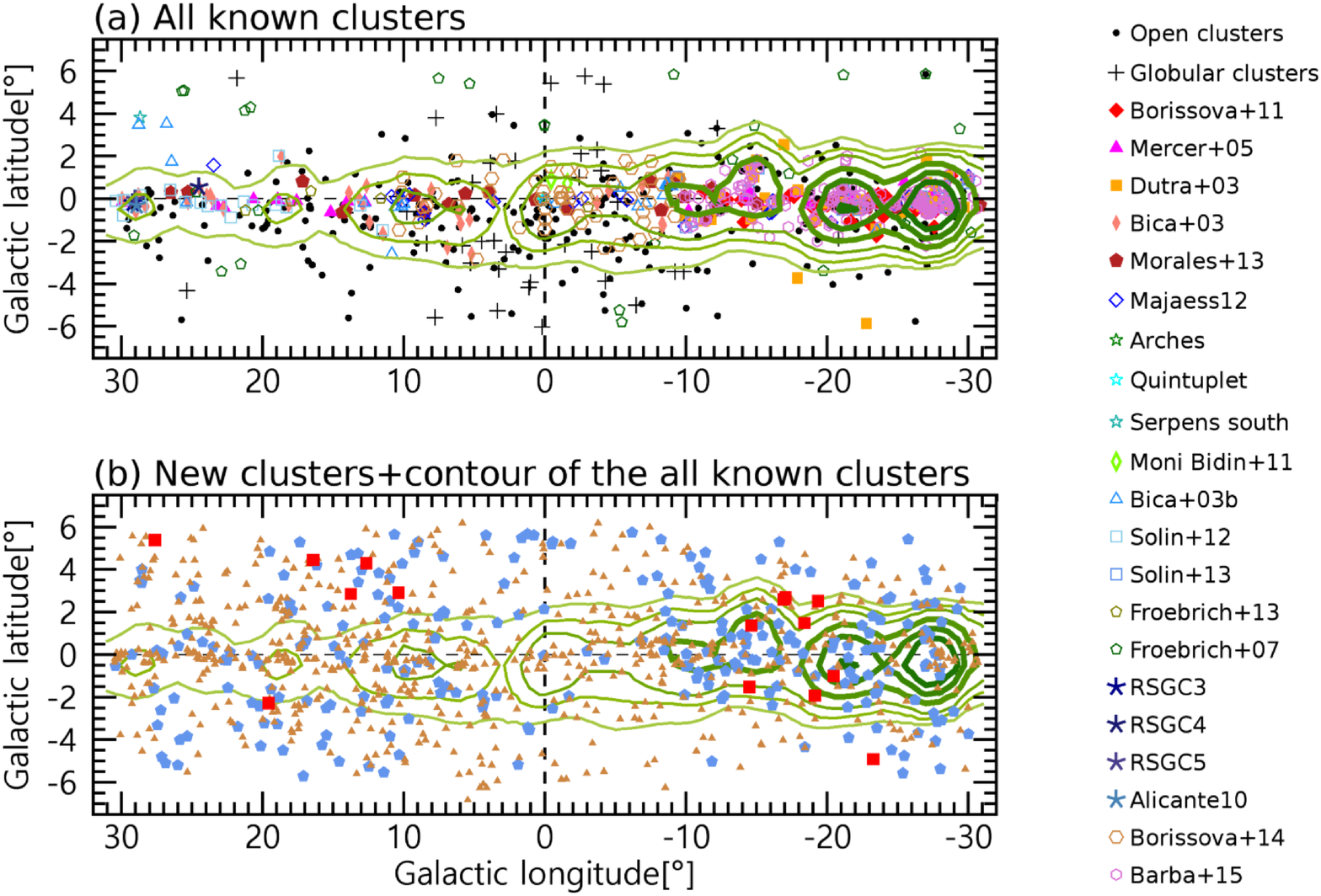}
\caption{(a) Spatial distribution of all known clusters located in the survey region. Corresponding reference or name of the cluster is listed on the right side. Contour levels represent 0$\sigma$, 0.5$\sigma$, 1$\sigma$, 2$\sigma$, 3$\sigma$, and 4$\sigma$ excess with respect to the average number density of the entire survey region. Thick contours are used to emphasize 2$\sigma$, 3$\sigma$, and 4$\sigma$ excess. (b) Spatial distribution of the new clusters. Filled squares, pentagons, and triangles represent class A, class B, and class C clusters, respectively. The contours are for the known clusters, same as in (a). \label{spd-known}}
\end{figure*}

We compare the spatial distributions of the new clusters with those of the known clusters in Figure \ref{spd-known}. The known clusters in the survey region are well concentrated onto the Galactic plane and are found more in the fourth Galactic quadrant, as shown in Figure \ref{spd-known}(a). In contrast, the new clusters are found more in the first Galactic quadrant, as shown in Figure \ref{spd-known}(b). This asymmetric distribution of the known clusters is a selection effect of previous studies. Before this study, only 19 clusters were found in our survey area from the UKIDSS data \citep{sol12}, while 320 clusters were found from the VVV data \citep{bor11, mon11, bor14, sol14, bar15}. There are four peaks in the number density in the distribution of the known clusters, at 
 $(l, b) \simeq$ ($-27\arcdeg.5$, $0\arcdeg$), ($-21\arcdeg.5$, $0\arcdeg$), ($-15\arcdeg$, $0\arcdeg.5$), and ($-10\arcdeg$, $0\arcdeg$). Among those peaks, the most prominent peak at $(l, b) \simeq$ ($-27\arcdeg.5$, $0\arcdeg$) coincides strongly with that of the new embedded clusters shown in Figure \ref{spd-cands}(b). This clumpy distribution is similar to the distribution of the new embedded clusters, which would be a common property of young clusters.

Another difference between the spatial distribution of the new clusters and that of the known clusters is shown in Figure \ref{spd-known}(b). Only a small number of clusters were discovered at $|b|>4\arcdeg$ regions because previous studies mostly focused on the Galactic plane. Clusters in that region are open clusters \citep{dia02}, globular clusters \citep{har96}, and the clusters found by \citet{fro07}. Most of the new clusters distributed in this high latitude region (except for the Aquila Rift region) may be comparable to the clusters found by \citet{fro07}, because of the spatial coincidence.

\begin{figure}
\epsscale{.85}
\plotone{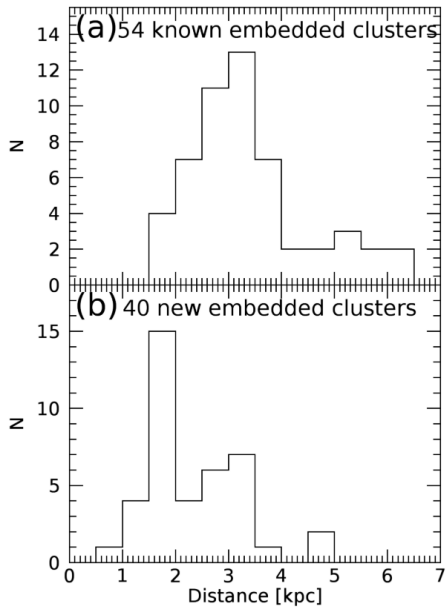}
\caption{(a) Distance distribution based on the size of known embedded clusters near the $(l, b) \simeq$ ($-27\arcdeg.5$, $0\arcdeg$), ($-21\arcdeg.5$, $0\arcdeg$), ($-15\arcdeg$, $0\arcdeg.5$), and ($-10\arcdeg$, $0\arcdeg$). (b) Distance distribution based on the size of the new embedded clusters near the $(l, b) \simeq$ ($-27\arcdeg.5$, $0\arcdeg$), ($-8\arcdeg$, $0\arcdeg.5$), ($17\arcdeg.5$, $1\arcdeg$), ($24\arcdeg$, $0\arcdeg$), and ($29\arcdeg$, $0\arcdeg$). \label{dpd}}
\end{figure}

If we assume that the typical radius of embedded clusters is 0.8 pc, which is the average radius of 36 known embedded clusters listed in \citet{lad03}, we can roughly estimate the distances to the known and new embedded clusters. We estimate the distances to the known embedded clusters near the four peaks based on their radii, and describe the distance distributions in Figure \ref{dpd}(a). Most of the known embedded clusters near the four peaks seem to be located at the Scutum-Centaurus arm ($\sim 3.1$ kpc) and the Norma arm ($\sim 4.6$ kpc). On the other hand, as shown in Figure \ref{dpd}(b), most of the new embedded clusters near the five peaks at $(l, b) \simeq$ ($-27\arcdeg.5$, $0\arcdeg$), ($-8\arcdeg$, $0\arcdeg.5$), ($17\arcdeg.5$, $1\arcdeg$), ($24\arcdeg$, $0\arcdeg$), and ($29\arcdeg$, $0\arcdeg$) seem to be located in the Sagittarius arm ($\sim 1.5$ kpc) and in the Scutum-Centaurus arm.

\subsection{Number distributions of the clusters in the Galactic latitude and longitude}
\begin{figure*}
\plotone{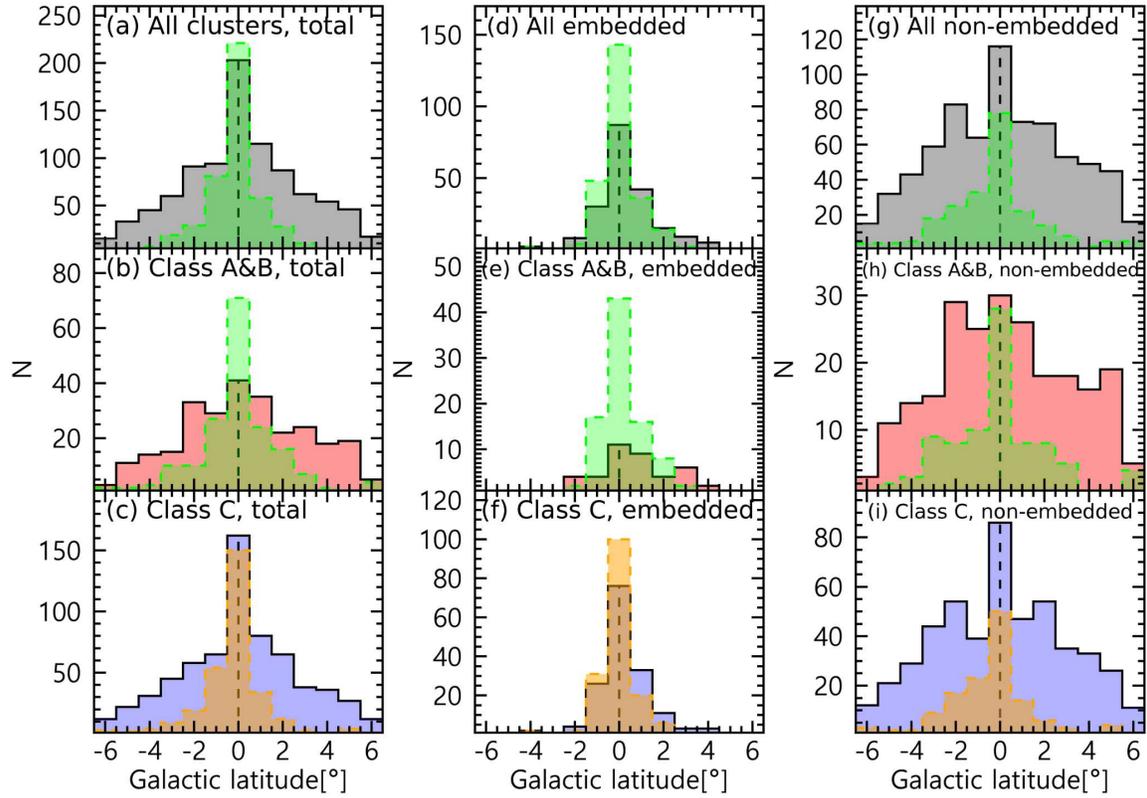}
\caption{Number distributions of the new clusters (solid line histograms) and known clusters (dashed line histograms) with respect to the Galactic latitude: (a) the entire set of new clusters, (b) the class A and B clusters, (c) the class C clusters, (d) all new embedded clusters, (e) embedded clusters in the class A and B, (f) embedded clusters in the class C, (g) all new non-embedded clusters, (h) non-embedded clusters in the class A and B, and (i) non-embedded clusters in the class C. \label{spdhist1}}
\end{figure*}

\begin{figure*}
\plotone{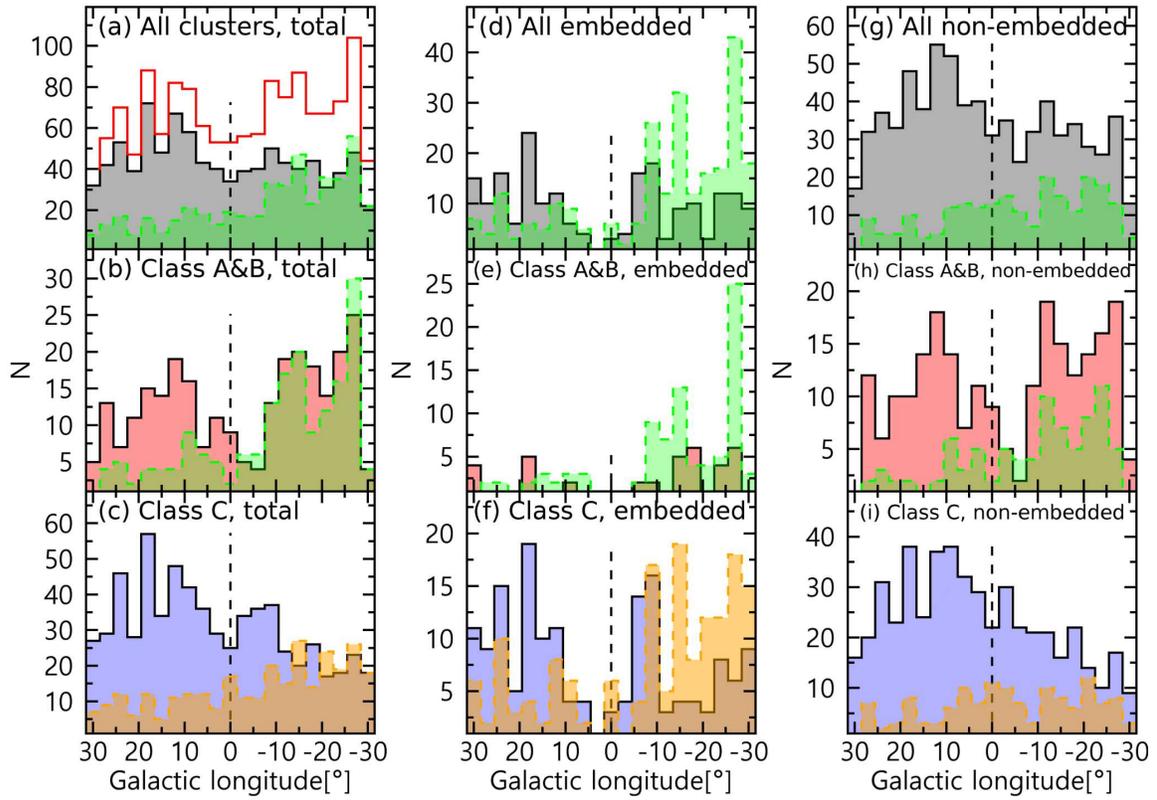}
\caption{Same as Figure \ref{spdhist1} but for with respect to the Galactic longitude. The empty solid line histogram in (a) represents the combined number distribution of the new clusters and the known clusters. \label{spdhist2}}
\end{figure*}

We plot the number distributions of the new clusters and known clusters with respect to the Galactic latitude and longitude in Figure \ref{spdhist1} and \ref{spdhist2}, respectively. For this comparison, we use 475 known clusters that satisfy our cluster selection criteria.

Symmetry is seen in the new and known clusters with respect to the Galactic latitude (Figure \ref{spdhist1}(a)(b)(c)). The symmetry can be seen in both the embedded clusters (Figure \ref{spdhist1}(d)) and the non-embedded clusters (Figure \ref{spdhist1}(g)). Also, all new clusters and the known clusters are concentrated and are similarly distributed onto the Galactic plane. For the embedded clusters, the new clusters and the known clusters are concentrated to the Galactic plane as shown in Figure \ref{spdhist1}(d) and (f), though the number of clusters and the survey area are different between the known clusters and the new clusters. Due to the small number of the new embedded clusters in class A+B, their concentration onto the Galactic plane is weaker than that of the known clusters (Figure \ref{spdhist1}(e)). On the other hand, for the non-embedded clusters, the new clusters are dispersed in the entire survey range of the Galactic latitude, while the known clusters are not, as shown in Figure \ref{spdhist1}(g), (h), and (i). This distribution difference between the new and known non-embedded clusters is due to the difference in field coverage: the survey area of this study is much larger than the previous infrared cluster surveys.


There is a weak asymmetry with respect to the Galactic longitude, as shown in Figure \ref{spdhist2}. For the new clusters, we find 101 more clusters in the first Galactic quadrant than in the fourth Galactic quadrant so that the number ratio of the new clusters between the two quadrants is 1.25 (=512:411). The class A+B clusters are found more in the fourth Galactic quadrant (Figure \ref{spdhist2}(b)(e)(h)), while the class C clusters are not (Figure \ref{spdhist2}(c)(f)(i)). In contrast, the strong asymmetry is seen in the known clusters, especially in embedded clusters (Figure \ref{spdhist2}(d)(e)(f)). The number ratio of the known clusters between the two quadrants, 0.46 (=150:325), is much different from that of the new clusters. However, the number ratio of the combined sample of new clusters and known clusters between the two quadrants is 0.90 (=662:736), as seen in Figure \ref{spdhist2}(a). Therefore we suggest that the number distribution of all star clusters along the Galactic plane is roughly symmetric.


\section{Summary}
The main goal of this study is to find new star clusters 
in the central plane region of the Milky Way. In the $|l|<30\arcdeg$ and $|b|<6\arcdeg$ region, we find 923 new clusters. They consist of 15 class A, 254 class B, and 654 class C clusters. Among them, 202 ($22\%$) clusters are classified as embedded clusters. In particular, $45\%$, $83\%$, and $55\%$ of the new clusters are strongly correlated with the IRAS point sources, the MSX sources, and the dark nebulae, respectively.

The average angular radius of the clusters is $1\arcmin.31\pm0\arcmin.60$. More specifically, 902 ($98\%$) clusters are smaller than 3$\arcmin$, and 823 ($89\%$) clusters are even smaller than 2$\arcmin$. It is due to the severe contamination from background stars that prevents finding of large and sparse clusters in the survey region.

We estimate reddenings, distances, and relative ages of the 15 class A clusters using the isochrone fitting. Reddenings and distances are determined at the $K_s$--$(H-K_s)$ CMD and the $K_s$--$(J-K_s)$ CMD, respectively. For these clusters, $E(B-V)$ values range from 0.2 to 5.2, and distances range from 440 pc to 12.6 kpc. They consist of 5 young clusters and 10 old clusters.

We investigate the spatial distribution of the new clusters. They show a clumpy structure while the entire set of known clusters in the survey region do not. The new clusters are concentrated onto the Galactic plane and distributed symmetrically with respect to the Galactic latitude. The new embedded clusters are more concentrated onto the Galactic plane than the new non-embedded clusters, which are dispersed throughout the entire survey range of the Galactic latitude.

On the other hand, the spatial distribution of the new clusters with respect to the Galactic longitude is asymmetric. The new clusters are populated more at the first Galactic quadrant, because the known clusters, especially the known embedded clusters are populated more at the fourth Galactic quadrant. The overall number ratio, which is counting all of the new clusters and known clusters is 0.90 (=662:736). This shows that the number distribution of all star clusters along the Galactic plane is roughly symmetric.


\acknowledgments
We thank the anonymous referee for several suggestions that have improved the quality of the manuscript. Also we thank Brian S. Cho for improving English in the manuscript.
This work was supported by the National Research Foundation of Korea (NRF) grant funded by the Korea Government (MSIP) (No. 2017R1A2B4004632). 
This research has made use of the NASA/IPAC Infrared Science Archive, which is operated by the Jet Propulsion Laboratory, California Institute of Technology, under contract with the National Aeronautics and Space Administration.


\begin{table*}
\begin{center}
\caption{A list of previous star cluster surveys overlapping with the field of this study.\label{t1}}
\begin{tabular}{lcrr}
\tableline\tableline
References & Used dataset & Total & N$_{cluster}$ included\\
& & N$_{cluster}$ & in this survey area\\
\tableline
\citet[globular clusters]{har96} & Compiled\tablenotemark{a} & 157 & 38\\
\citet[open clusters]{dia02} & Compiled\tablenotemark{a} & 2174 & 244\smallskip\\
\citet{nag90} & IRTF(NIR)\tablenotemark{b} & 1 & 1\\	
\citet{nag95} & IRTF(NIR)\tablenotemark{b} & 1 & 1\\
\citet{bic03a} & Compiled\tablenotemark{a} & 276 & 23\\ 
\citet{bic03b} & 2MASS(NIR) & 167 & 41\\
\citet{dut03} & 2MASS(NIR) & 179 & 55\\
\citet{mer05} & GLIMPSE(MIR) & 92 & 32\\
\citet{fro07} & 2MASS(NIR) & 1021 & 28\\ 
\citet{gut08} & \textit{Spitzer}(MIR)\tablenotemark{c} & 1 & 1\\
\citet{ale09} & GLIMPSE(MIR) & 1 & 1\\
\citet{neg10} & UKIDSS(NIR) & 1 & 1\\
\citet{bor11} & VVV(NIR) & 96 & 41\\
\citet{mon11} & VVV(NIR) & 3 & 2\\
\citet{neg11} & 2MASS(NIR) & 1 & 1\\
Gonz\'alez-Fern\'andez & 2MASS(NIR) & 1 & 1\\
$\quad$\& Negueruela (2012) & & &\\
\citet{sol12} & UKIDSS(NIR) & 137 & 19\\
\citet{fro13} & NIR surveys\tablenotemark{d} & 6 & 5\\
\citet{maj13} & WISE(MIR) & 229 & 21\\
\citet{mor13} & GLIMPSE(MIR) & 75 & 27\\
\citet{bor14} & VVV(NIR) & 58 & 58\\
\citet{sol14} & VVV(NIR) & 88 & 21\\
\citet{bar15} & VVV(NIR) & 493 & 198\\
\citet{cam16} & WISE(MIR) & 1098\tablenotemark{e} & 37\\
\tableline
\end{tabular}
\end{center}
\tablenotetext{a}{The catalog was compiled using various individual references.}
\tablenotetext{b}{It is a targeting observation using the NASA Infrared Telescope Facility.}
\tablenotetext{c}{It is part of the \textit{Spitzer} Gould Belt Legacy Survey.}
\tablenotetext{d}{The catalog was made by using the 2MASS, UKIDSS, and VVV data.}
\tablenotetext{e}{This number of clusters includes 437, 2, and 7 clusters found in \citet{cam15a}, \citet{cam15b}, and \citet{cam15c}, respectively.}
\end{table*}

\begin{table*}
\small
\begin{center}
\caption{Grading conditions for cluster candidates.\label{t2}}
\begin{tabular}{ccl}
\tableline\tableline
Criterion & Point &$\quad$ Conditions\\
\tableline
Morphology & 2 & $I_{W1}(0) \geq I_{W1,bg}+2\sigma_{bg}$\\ \cline{2-3}
& 1 & $I_{W1}(0) \leq I_{W1,bg}-2\sigma_{bg}$ OR\\
& & ($I_{W1,bg}-2\sigma_{bg} < I_{W1}(0) < I_{W1,bg}+2\sigma_{bg} \ AND\ f_{fs}\tablenotemark{a} \geq 0.32$)\\ \cline{2-3}
& 0 & $I_{W1,bg}-2\sigma_{bg}<I_{W1}(0)<I_{W1,bg}+2\sigma_{bg} \ AND\  f_{fs}<0.32$\\
\tableline
CMD & 2 & ($N_{star\_cluster} \geq 50$ AND $f_{remaining\_star}\tablenotemark{b}>0.72\ (\textrm{without deep NIR})$) OR\\
& & ($N_{star\_cluster} \geq 50$ AND $f_{remaining\_star}>0.51\ (\textrm{with deep NIR})$)\\ \cline{2-3}
& 1 & ($N_{star\_cluster} \geq 50 \ AND\ 0.1 \leq f_{remaining\_star} \leq 0.72\ (\textrm{without deep NIR})$) OR\\
& & ($N_{star\_cluster} \geq 50 \ AND\ 0.1 \leq f_{remaining\_star} \leq 0.51\ (\textrm{with deep NIR})$)\\ \cline{2-3}
& 0 & $N_{star\_cluster}<50 \ OR \ f_{remaining\_star}<0.1$\\
\tableline
RDP & 2 & $\bullet$ Central number density excesses ($n_{peak}(r<R_{cl}) > n_{bg}+2\sigma_{star\_bg}$) in both the $K_s$ and the W1 band\\ \cline{2-3}
& 1 & $\bullet$ A central number density excess in one band\\ \cline{2-3}
& 0 & $\bullet$ No central number density excess in any band\\
\tableline
\end{tabular}
\end{center}
\tablenotetext{a}{A spatial fraction of unresolved stellar light inside the cluster region. We regard signals with $I_{W1,bg}\leq I_{W1}\leq I_{W1,bg}+2\sigma_{bg}$ as unresolved stellar light.}
\tablenotetext{b}{A number fraction of remaining stars after the background subtraction with respect to the total number of stars in the CMD of the cluster region ($N_{star\_cluster}$).}
\end{table*}

\begin{table*}
\begin{center}
\small
\caption{A list of new clusters.\label{t3}}
\begin{tabular}{cccccccc}
\tableline\tableline
Name & \it{l} & \it{b} & RA(J2000) & Dec(J2000) & Radius & Embedded & Class\\
 & (deg) & (deg) & (hh:mm:ss) & (dd:mm:ss) & (arcmin)\\
\tableline
Ryu 001 & 336.7381 & --4.9119 & 16:57:20.37 & --50:52:52.7 & 1.6 & $\times$ & A\\
Ryu 002 & 339.5574 & --1.0115 & 16:49:46.45 & --46:14:11.3 & 1.2 & $\times$ & A\\
Ryu 003 & 340.6465 & 2.5251 & 16:38:43.69 & --43:05:58.4 & 1.0 & $\times$ & A\\
Ryu 004 & 340.8815 & --1.9155 & 16:58:37.27 & --45:46:43.5 & 1.6 & $\bigcirc$ & A\\
Ryu 005 & 341.6239 & 1.4827 & 16:46:32.72 & --43:02:55.6 & 0.9 & $\times$ & A\\
\tableline
\end{tabular}
\end{center}
\tablecomments{The full version of the catalog is available online.}
\end{table*}

\begin{deluxetable*}{cccccc}
\tabletypesize{\small}
\tablecaption{A list of the sources associated with new clusters.\label{t4}}
\tablewidth{0pt}
\tablehead{
\colhead{Name} & \colhead{IRAS PSC} & \colhead{MSX6C} & \colhead{Dark nebulae} & \colhead{YSO} & \colhead{Masers}
}
\startdata
Ryu 001 & 16533--5048 & G336.7244--04.9023 & - & - & -\\
Ryu 002 & 16461--4608 & G339.5593--01.0123 & - & AKARI 1649472--46140 & -\\
Ryu 003 & - & G340.6485+02.5251 & FeSt1--356 & - & -\\
Ryu 004 & 16551--4541 & G340.8683--01.9268 & - & AKARI 1658376--45472 & -\\
Ryu 005 & 16430--4257 & G341.6248+01.4821 & - & - & -\\
\enddata
\tablecomments{The full version of the catalog is available online.}
\end{deluxetable*}

\begin{table*}
\begin{center}
\caption{A cross-matching result for the IRAS point sources, MSX sources, and dark nebulae.\label{t5}}
\begin{tabular}{ccccc}
\tableline\tableline
Subset & Class A & Class B & Class C & Total\\
\tableline
\small New clusters & 15 & 254 & 654 & 923\\
\small Embedded clusters & \small5 ($33\%$; 5/15) & \small35 ($14\%$; 35/254) & \small162 ($25\%$; 162/654) & \small202 ($22\%$; 202/923)\\
\tableline
\small IRAS point sources & \small10 ($67\%$; 10/15) & \small108 ($43\%$; 108/254) & \small294 ($45\%$; 294/654) & \small412 ($45\%$; 412/923)\\
\small Embedded clusters & \small4 ($80\%$; 4/5) & \small26 ($74\%$; 26/35) & \small126 ($78\%$; 126/162) & \small156 ($77\%$; 156/202)\\
\tableline
\small MSX sources & \small14 ($93\%$; 14/15) & \small216 ($85\%$; 216/254) & \small537 ($82\%$; 537/654) & \small767 ($83\%$; 767/923)\\
\small Embedded clusters & \small5 ($100\%$; 5/5) & \small34 ($97\%$; 34/35) & \small157 ($97\%$; 157/162) & \small196 ($97\%$; 196/202)\\
\tableline
\small Dark nebulae & \small10 ($67\%$; 10/15) & \small129 ($51\%$; 129/254) & \small368 ($56\%$; 368/654) & \small507 ($55\%$; 507/923)\\
\small Embedded clusters & \small4 ($80\%$; 4/5) & \small30 ($86\%$; 30/35) & \small133 ($82\%$; 133/162) & \small167 ($83\%$; 167/202)\\
\tableline
\end{tabular}
\end{center}
\end{table*}

\begin{table*}
\small
\begin{center}
\caption{Reddenings, distances, and relative ages of class A clusters.\label{t6}}
\begin{tabular}{cccccccccc}
\tableline\tableline
Name & \it{l} & \it{b} & RA(J2000) & Dec(J2000) & Radius & Embedded & $E(B-V)$ & Distance & Age\\
 & (deg) & (deg) & (hh:mm:ss) & (dd:mm:ss) & (arcmin) & & (mag) & (kpc)\\
\tableline
Ryu 001 & 336.7381 & --4.9119 & 16:57:20.37 & --50:52:52.7 & 1.6 & $\times$ & $0.2\pm0.1$ & $6.3^{+3.7}_{-2.3}$ & Old\\
Ryu 002 & 339.5574 & --1.0115 & 16:49:46.45 & --46:14:11.3 & 1.2 & $\times$ & $0.9\pm0.4$ & $4.6^{+1.2}_{-1.0}$ & Old\tablenotemark{a}\\
Ryu 003 & 340.6465 & 2.5251 & 16:38:43.69 & --43:05:58.4 & 1.0 & $\times$ & $1.8\pm0.2$ & $7.9^{+2.1}_{-1.6}$ & Old\\
Ryu 004 & 340.8815 & --1.9155 & 16:58:37.27 & --45:46:43.5 & 1.6 & $\bigcirc$ & $3.1\pm0.2$ & $0.5^{+0.3}_{-0.2}$ &Young\\
Ryu 005 & 341.6239 & 1.4827 & 16:46:32.72 & --43:02:55.6 & 0.9 & $\times$ & $2.2\pm0.2$ & $7.2^{+1.9}_{-1.4}$ & Old\\
Ryu 006 & 342.9710 & 2.6796 & 16:46:20.24 & --41:14:52.3 & 1.5 & $\bigcirc$ & $2.2\pm0.3$ & $0.5^{+0.2}_{-0.1}$ & Young\\
Ryu 007 & 343.0815 & 2.5970 & 16:47:03.15 & --41:13:02.0 & 1.9 & $\bigcirc$ & $3.3^{+0.3}_{-1.0}$ & $0.5^{+0.2}_{-0.1}$ & Young\\
Ryu 008 & 345.3693 & 1.3649 & 16:59:43.16 & --40:13:26.4 & 1.4 & $\bigcirc$ & $5.2^{+0.2}_{-0.4}$ & $0.4\pm0.1$ & Young\\
Ryu 009 & 345.5155 & --1.5256 & 17:12:24.85 & --41:51:04.1 & 1.0 & $\times$ & $0.9^{+0.2}_{-0.3}$ & $12.6^{+4.8}_{-3.5}$ & Old\\
Ryu 010 & 10.3712 & 2.9127 & 17:57:47.15 & --18:31:57.3 & 2.3 & $\bigcirc$ & $3.8\pm0.4$ & $0.5^{+0.3}_{-0.2}$ & Young\\
Ryu 011 & 12.6436 & 4.2902 & 17:57:30.59 & --15:52:48.0 & 1.1 & $\times$ & $1.0\pm0.1$ & $2.9\pm0.4$ & Old\\
Ryu 012 & 13.7591 & 2.8622 & 18:04:57.19 & --15:36:50.4 & 1.2 & $\times$ & $0.8\pm0.1$ & $6.9^{+1.0}_{-0.9}$ & Old\\
Ryu 013 & 16.4291 & 4.4530 & 18:04:34.47 & --12:30:43.7 & 1.5 & $\times$ & $1.1\pm0.2$ & $4.4^{+1.6}_{-1.2}$ & Old\tablenotemark{b}\\
Ryu 014 & 19.5743 & --2.2773 & 18:34:59.35 & --12:55:07.4 & 1.1 & $\times$ & $0.5^{+0.1}_{-0.2}$ & $5.0^{+1.3}_{-1.0}$ & Old\tablenotemark{a}\\
Ryu 015 & 27.6273 & 5.3777 & 18:22:36.30 & --02:14:44.2 & 2.7 & $\times$ & $1.1\pm0.1$ & $3.6^{+2.7}_{-1.5}$& Old\tablenotemark{b}\\
\tableline
\end{tabular}
\end{center}
\tablecomments{$E(B-V)$ and distances of young and old clusters are determined using 4 Myr and 10 Gyr isochrones, respectively.}
\tablenotetext{a}{The age is determined to be $0.8\pm0.2$ Gyr.}
\tablenotetext{b}{The age is determined to be $2.0^{+0.5}_{-0.4}$ Gyr.}
\end{table*}



\clearpage

\clearpage


\appendix
\renewcommand\thefigure{\thesection\arabic{figure}}
\setcounter{figure}{0}
\section{CMDs, gray-scale maps, and RDP of 15 class A clusters}
\begin{figure*}
\plotone{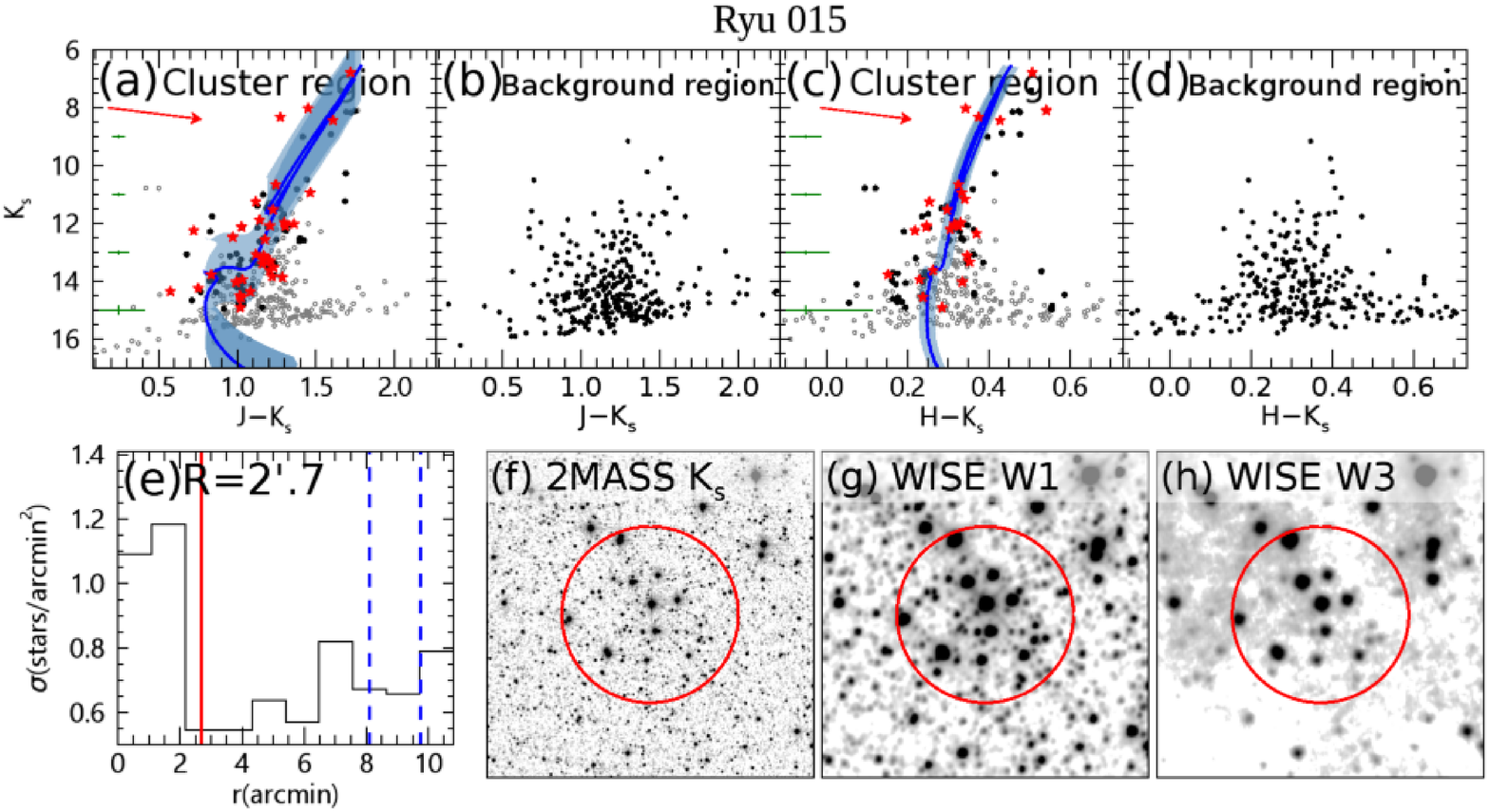}
\caption{Figures for an example of the new cluster, Ryu 015. (a) $K_s$--$(J-K_s)$ CMD of the cluster region. Red starlets and black filled dots represent stars located at $r\leq0.5R_{cl}$ and $0.5R_{cl}<r\leq R_{cl}$, respectively. Gray open circles represent stars removed by the statistical background subtraction process. A red arrow represents the direction of the reddening vector. A blue line is the best-fit isochrone, and the shaded region around the line represents an error of the distance modulus estimation. Typical errors of magnitude and color are represented by error bars at the left side of the CMD. (b) $K_s$--$(J-K_s)$ CMD of the background region. (c) and (d) $K_s$--$(H-K_s)$ CMDs of the cluster and the background region, respectively. Symbols are the same as in (a). The shaded region in this CMD represents an error of the reddening estimation. (e) An RDP of the cluster derived from the W1 band data. A vertical red solid line represents the size of the cluster, and two blue dashed lines represent the inner ($3R_{cl}$) and outer ($3.6R_{cl}$) radii of the background region. (f), (g), and (h) The $K_s$, W1, and W3 band gray-scale images of the cluster, respectively. The field of view of each image is $10\arcmin\times10\arcmin$. Circles represent the size of the cluster. North is up, and east is left. \label{figa1}}
\end{figure*}

We present CMDs, an RDP, and gray-scale maps of Ryu 015 in Figure \ref{figa1} as an example. A complete figure set for the class A clusters is available in the online journal.

\figsetstart
\figsetnum{A1}
\figsettitle{CMDs, gray-scale maps, and RDP of 15 class A clusters}

\figsetgrpstart
\figsetgrpnum{A1.1}
\figsetgrptitle{Ryu 001}
\figsetplot{ryu001fig.eps}
\figsetgrpnote{Figures of Ryu 001. Symbols are same as Figure \ref{figa1}.}
\figsetgrpend

\figsetgrpstart
\figsetgrpnum{A1.2}
\figsetgrptitle{Ryu 002}
\figsetplot{ryu002fig.eps}
\figsetgrpnote{Figures of Ryu 002. Symbols are same as Figure \ref{figa1}. (e) The RDP of this cluster is made using the GLIMPSE [3.6] sources, because the spatial resolution of the GLIMPSE [3.6] band is better than that of the WISE W1 band.}
\figsetgrpend

\figsetgrpstart
\figsetgrpnum{A1.3}
\figsetgrptitle{Ryu 003}
\figsetplot{ryu003fig.eps}
\figsetgrpnote{Figures of Ryu 003. Symbols are same as Figure \ref{figa1}.}
\figsetgrpend

\figsetgrpstart
\figsetgrpnum{A1.4}
\figsetgrptitle{Ryu 004}
\figsetplot{ryu004fig.eps}
\figsetgrpnote{Figures of Ryu 004. Symbols are same as Figure \ref{figa1}.}
\figsetgrpend

\figsetgrpstart
\figsetgrpnum{A1.5}
\figsetgrptitle{Ryu 005}
\figsetplot{ryu005fig.eps}
\figsetgrpnote{Figures of Ryu 005. Symbols are same as Figure \ref{figa1}. (e) The RDP of this cluster is made using the GLIMPSE [3.6] sources, because the spatial resolution of the GLIMPSE [3.6] band is better than that of the WISE W1 band.}
\figsetgrpend

\figsetgrpstart
\figsetgrpnum{A1.6}
\figsetgrptitle{Ryu 006}
\figsetplot{ryu006fig.eps}
\figsetgrpnote{Figures of Ryu 006. Symbols are same as Figure \ref{figa1}.}
\figsetgrpend

\figsetgrpstart
\figsetgrpnum{A1.7}
\figsetgrptitle{Ryu 007}
\figsetplot{ryu007fig.eps}
\figsetgrpnote{Figures of Ryu 007. Symbols are same as Figure \ref{figa1}.}
\figsetgrpend

\figsetgrpstart
\figsetgrpnum{A1.8}
\figsetgrptitle{Ryu 008}
\figsetplot{ryu008fig.eps}
\figsetgrpnote{Figures of Ryu 008. Symbols are same as Figure \ref{figa1}.}
\figsetgrpend

\figsetgrpstart
\figsetgrpnum{A1.9}
\figsetgrptitle{Ryu 009}
\figsetplot{ryu009fig.eps}
\figsetgrpnote{Figures of Ryu 009. Symbols are same as Figure \ref{figa1}.}
\figsetgrpend

\figsetgrpstart
\figsetgrpnum{A1.10}
\figsetgrptitle{Ryu 010}
\figsetplot{ryu010fig.eps}
\figsetgrpnote{Figures of Ryu 010. Symbols are same as Figure \ref{figa1}. (e) The RDP of this cluster is made using the GLIMPSE [3.6] sources, because the spatial resolution of the GLIMPSE [3.6] band is better than that of the WISE W1 band.}
\figsetgrpend

\figsetgrpstart
\figsetgrpnum{A1.11}
\figsetgrptitle{Ryu 011}
\figsetplot{ryu011fig.eps}
\figsetgrpnote{Figures of Ryu 011. Symbols are same as Figure \ref{figa1}.}
\figsetgrpend

\figsetgrpstart
\figsetgrpnum{A1.12}
\figsetgrptitle{Ryu 012}
\figsetplot{ryu012fig.eps}
\figsetgrpnote{Figures of Ryu 012. Symbols are same as Figure \ref{figa1}.}
\figsetgrpend

\figsetgrpstart
\figsetgrpnum{A1.13}
\figsetgrptitle{Ryu 013}
\figsetplot{ryu013fig.eps}
\figsetgrpnote{Figures of Ryu 013. Symbols are same as Figure \ref{figa1}.}
\figsetgrpend

\figsetgrpstart
\figsetgrpnum{A1.14}
\figsetgrptitle{Ryu 014}
\figsetplot{ryu014fig.eps}
\figsetgrpnote{Figures of Ryu 014. Symbols are same as Figure \ref{figa1}. (e) The RDP of this cluster is made using the GLIMPSE [3.6] sources, because the spatial resolution of the GLIMPSE [3.6] band is better than that of the WISE W1 band.}
\figsetgrpend

\figsetgrpstart
\figsetgrpnum{A1.15}
\figsetgrptitle{Ryu 015}
\figsetplot{ryu015fig.eps}
\figsetgrpnote{Figures of Ryu 015. Symbols are same as Figure \ref{figa1}.}
\figsetgrpend

\figsetend

\section{The statistical CMD background subtraction method}
The stellar sequence of the cluster 
in the Galactic plane region is usually contaminated by the background stars so that the subtraction method is especially useful for the clusters we found.
The statistical background subtraction from the CMD of 
a cluster is based on the idea that 
the member stars of a cluster are located at the different region in the CMD from the background field stars in the CMD.

For each member star of the cluster, we set a magnitude range and a color range to compare the cluster sample with the background sample. The magnitude range is $\pm0.5$ mag from the target star. The color range depends on the standard deviation of the color distribution of stars in the CMD of the cluster region; we use $\pm0.3\sigma$ for the color range of the target star. Within these ranges, we count the numbers of the stars in the cluster region and the stars in the background region. During this counting process, we make a number of subsets of stars in each region for the bootstrap resampling. We also consider the area ratio: 
the area of our background region is 4 times wider than that of the cluster region.

We define the survival probability of each member star of the cluster using 
star counts within its magnitude and color ranges. The probability 
depends not only on the target star but also 
on its neighbor stars. 
We perform $N_{boot}$ times of the bootstrap resampling. Therefore, the survival probability is calculated by the following equation: \begin{equation} P_{surv\_target}=\frac{\sum_{}^{N_{boot}}(N_{star\_cluster} - N_{star\_background})}{N_{boot}\times(N_{star\_cluster})}
,
 \end{equation}
\noindent
where $N_{star\_cluster}$ and $N_{star\_background}$ are the numbers of the stars within the ranges of the cluster and the background region, respectively. We consider the star is survived from the background subtraction if 
its survival probability 
is over $50\%$.



\end{document}